\documentclass[journal]{IEEEtran}
\ifCLASSINFOpdf
  \usepackage[pdftex]{graphicx}
\else
\fi
\usepackage{amsmath}
\usepackage{array}

\usepackage{times}
\usepackage{epsfig}
\usepackage{amssymb}
\usepackage{array}
\usepackage{multirow}
\usepackage{threeparttable}
\usepackage{color}
\newcommand{\wgc}[1]{\textcolor{black}{#1}}
\usepackage{url}

\begin{document}
%
\title{Spatial-Angular Attention Network for\\Light Field Reconstruction}
%
%
%

\author{Gaochang~Wu,~
		Yingqian~Wang,~
		Yebin~Liu,~\IEEEmembership{Member,~IEEE,}
				Lu~Fang,~\IEEEmembership{Senior Member,~IEEE,}
        and~Tianyou~Chai,~\IEEEmembership{Fellow,~IEEE}
\thanks{This work was supported by the Major Program of National Natural Science Foundation of China No.61991400, No.61991401 and No.62103092, Natural Science Foundation of China No. U20A20189, Science and Technology Major Projects of Liaoning Province No.2020JH1/10100008, NSFC No.61827805, No.61531014, No.61861166002 and No.6181001011, and Fundamental Research Funds for the Central Universities No. 100802004. (\textit{Corresponding author: Tianyou Chai}.)}
\thanks{Gaochang Wu and Tianyou Chai are with the State Key Laboratory of Synthetical Automation for Process Industries, Northeastern University, Shenyang 110819, China, and also with the Institute of Industrial Artificial Intelligence, Northeastern University, Shenyang 110819, P. R. China (email: \{wugc, tychai\}@mail.neu.edu.cn).}
\thanks{Yingqian Wang is with the College of Electronic Science and Technology, Nation University of Defense Technology (NUDT), P. R. China. (email: wangyingqian16@nudt.edu.cn).}
\thanks{Yebin Liu is with Department of Automation, Tsinghua University, Beijing 100084, P. R. China (email: liuyebin@mail.tsinghua.edu.cn).}
\thanks{Fang Lu is with the Department of Electronic Engineering, Tsinghua University, Beijing 100084, China, and also with the Beijing National Research Center for Information Science and Technology, Beijing 100084, P. R. China (email: fanglu@tsinghua.edu.cn).}
}

%
%

\markboth{Journal of \LaTeX\ Class Files,~Vol.~14, No.~8, August~2015}%
{Wu \MakeLowercase{\textit{et al.}}: Bare Demo of IEEEtran.cls for IEEE Journals}
%



\maketitle

\begin{abstract}
Typical learning-based light field reconstruction methods demand in constructing a large receptive field by deepening their networks to capture correspondences between input views. In this paper, we propose a spatial-angular attention network to perceive non-local correspondences in the light field, and reconstruct high angular resolution light field in an end-to-end manner. Motivated by the non-local attention mechanism~\cite{wang2018non-local,zhang2018self}, a spatial-angular attention module specifically for the high-dimensional light field data is introduced to compute the response of each query pixel from all the positions on the epipolar plane, and generate an attention map that captures correspondences along the angular dimension. Then a multi-scale reconstruction structure is proposed to efficiently implement the non-local attention in the low resolution feature space, while also preserving the high frequency components in the high-resolution feature space. Extensive experiments demonstrate the superior performance of the proposed spatial-angular attention network for reconstructing sparsely-sampled light fields with non-Lambertian effects.
\end{abstract}


%
\IEEEpeerreviewmaketitle

\section{Introduction}
%
%
%
%
\IEEEPARstart{T}{hrough} capturing both intensities and directions from sampled light rays, light field enables high-quality view synthesis without the need of complex and heterogeneous information such as geometry and texture. More importantly, benefiting from the light field rendering technology~\cite{LFrendering}, photorealistic views can be rendered in real-time regardless of the scene complexity or non-Lambertian effect. This high quality rendering technology usually requires a densely-sampled light field (DSLF), where the disparity between adjacent views should be less than one pixel. However, typical DSLF capture either suffers from a long period of acquisition time (e.g., DSLF gantry system~\cite{LFrendering}) or falls into the well-known resolution trade-off problem. That is, due to the limitation of the sensor resolution~\cite{Lytro}, the light field is sparsely sampled either in the angular domain~\cite{LFrig} or the spatial domain~\cite{wang2020lfinternet}.

Recently, a more promising way is the fast capturing of a sparsely-sampled (angular domain) light field followed by direct reconstruction or depth-based view synthesis methods~\cite{DoubleCNN,shi2020learning} by using advanced deep learning techniques. On the one hand, typical learning-based reconstruction methods~\cite{LFCNN,WuEPICNN2018,YeungECCV2018} employ multiple convolutional layers to map the low angular resolution light field to the DSLF. But due to the limited perceptive range of convolutional filters~\cite{Long2014Do}, these networks fail to collect enough information (i.e., the spatial-angular correspondences) when dealing with large disparities, leading to aliasing effects in the reconstructed light field. On the other hand, depth-based view synthesis methods~\cite{DoubleCNN,DeepStereo,mildenhall2019local} address the large disparity problem through plane sweep (depth estimation), and then synthesize novel views using learning-based prediction. However, these methods require depth consistency along the angular dimension, and thus, often fail to handle the depth ambiguity caused by the non-Lambertian effect.

\begin{figure}
\begin{center}
\includegraphics[width=1\linewidth]{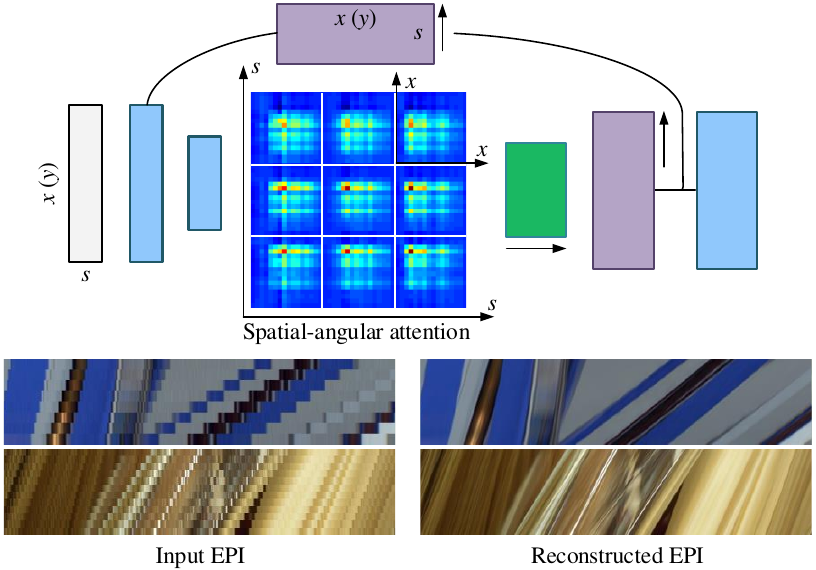}
\end{center}
\vspace{-4mm}
   \caption{We propose a spatial-angular attention module embedded in a multi-scale reconstruction structure for learning-based light field reconstruction. The network perceives correspondence pixels in a non-local manner, providing high quality reconstruction with a sparse input. Light fields courtesy of Moreschini~\textit{et al.}~\cite{ICME2018} and Adhikarla~\textit{et al.}~\cite{kiran2017towards}.}
\label{fig:Teaser}
\vspace{-2mm}
\end{figure}

In this paper, we propose a Spatial-Angular Attention Network, termed as SAA-Net, to achieve DSLF reconstruction from a sparse input. The proposed SAA-Net perceives correspondences on the Epipolar Plane Image (EPI) in a non-local fashion, addressing the aforementioned non-Lambertian issue and large disparity issue in a unified framework. Specifically, the SAA-Net consists of two parts, including a spatial-angular attention module (Sec. \ref{Sec:Network_archi}) and a U-net backbone (Sec. \ref{Sec:SAAM}). Motivated by the non-local attention mechanism in~\cite{wang2018non-local, zhang2018self}, for each pixel in the input light field, the Spatial-Angular Attention Module (termed as SAAM for short) computes the responses of pixels from all the positions on the epipolar plane, and generates an attention map that records the correspondences along the angular dimension, as shown in Fig. \ref{fig:Teaser} (top). This correspondence information in the attention map is then applied to guide the reconstruction in the angular dimension via matrix multiplication and channel-to-angular pixel shuffling.

To efficiently perform the non-local attention, we propose a convolutional neural network with multi-scale reconstruction structure. The network follows the basic architecture of the U-net, i.e., an encoder-decoder structure with skip connections. The encoder compresses the input light field in the spatial dimensions and removes redundancy information for the SAAM. Rather than simply reconstruct the light field at the end of the network, we propose a multi-scale reconstruction structure by performing deconvolution along the angular dimension in each skip connection branch, as shown in Fig. \ref{fig:Teaser} (top). The proposed multi-scale reconstruction structure maintains the view consistency in the low spatial resolution feature space while preserving fine details in the high spatial resolution feature space.

For network training, we propose a spatial-angular perceptual loss that is specifically designed for the high-dimensional light field data (Sec. \ref{Sec:training}). Rather than computing the high-level feature loss~\cite{dosovitskiy2016generating, Johnson2016Perceptual} by feeding each view in the light field into a 2D CNN (e.g., the commonly-used VGG~\cite{simonyan2015very}), we pre-train a 3D auto-encoder that considers the consistency in both the spatial and angular dimensions of the light field. In summary, we make the following  contributions\footnote{The source code is available at \url{https://github.com/GaochangWu/SAAN}.}:
\begin{itemize}
    \item A spatial-angular attention module that perceives correspondences non-locally on the epipolar plane;
    \item A multi-scale reconstruction structure for efficiently performing the non-local attention in the low spatial resolution feature space while also preserving high frequencies;
    \item A spatial-angular perceptual loss specifically designed for high-dimensional light field data.
\end{itemize}

We demonstrate the superiority of the  SAA-Net by performing extensive evaluations on various light field datasets. The proposed network presents high-quality DSLF on challenging cases with both non-Lambertian effects and large disparities, as illustrated in Fig. \ref{fig:Teaser} (bottom).

\section{Related Work}
\subsection{Light Lield Reconstruction}
First, we briefly review the major works on light field view synthesis (or view synthesis) depending on whether the depth information is explicitly used.

\textbf{Depth image-based view synthesis.} \wgc{Typically, these kind of approaches first estimate the depth of a scene, then warp and blend the input views to synthesize a novel view~\cite{soft3D,shi2020learning}}. Conventional light field depth estimation approaches follow the pipeline of stereo matching~\cite{scharstein2002taxonomy}, i.e., cost computation, cost aggregation (or cost volume filtering), disparity regression and post refinement. The main difference is that light field converts disparity from discrete space into a continuous space~\cite{Wanner}, delivering various depth cues that enable light field depth estimation with different strategies, such as structure tensor-based local direction estimation~\cite{Wanner}, \wgc{depth from correspondence~\cite{huang2017robust}, depth from defocus~\cite{Tao,Occ} and depth from parallelogram cues~\cite{zhang2016robust}.} Moreover, some learning-based approaches incorporate the aforementioned depth estimation pipeline with 2D convolution-based feature extraction, 3D convolution-based cost volume aggregation and depth regression~\cite{kendall2017end}. For novel view synthesis, input views are warped to the novel viewpoints with sub-pixel accuracy using bilinear interpolation and blended in different manners, e.g., total variation optimization~\cite{Wanner}, soft blending~\cite{soft3D} and learning-based synthesis~\cite{Zheng2018ECCV}.

Recently, researchers mainly focus on the studies for maximizing the quality of synthesized views based on the deep learning technique. Flynn \textit{et al.}~\cite{DeepStereo} proposed a learning-based method to synthesize novel views using the predicted probabilities and colors for each depth plane. Kalantari \textit{et al.}~\cite{DoubleCNN} further employed a sequential network setting to infer depth (disparity) and color, and optimized the model via end-to-end training. \wgc{Following the sequential network setting, Meng~\textit{et al.}~\cite{meng2021light} developed a confidence estimation network between depth and color networks to infer pixel-wise blending weights. Shi \textit{et al.}~\cite{shi2020learning} proposed to blend the warped views in both pixel level and feature level. Jin \textit{et al.} proposed to use a regular sampling pattern (four corner views)~\cite{jin2020learning} and a flexible sampling pattern~\cite{jin2020deep} for light field depth estimation, and then perform the reconstruction using spatial-angular alternating refinement. Ko~\textit{et al.}~\cite{ko2021light} introduced a dynamic blending filter to generate the filter coefficients adaptively according to the warped views. Different from the sequential network settings mentioned above,} Zhou \textit{et al.}~\cite{zhou2018stereo} proposed a novel learning-based Multi-Plane Image (MPI) representation that infers a novel view by alpha blending of different images. Mildenhall \textit{et al.}~\cite{mildenhall2019local} further proposed to use multiple MPIs to synthesize a local light field.

\wgc{Depth image-based view synthesis approaches solve the problem of large correspondence gap in the sparsely-sampled light field by using depth estimation and warping. But the scene depth are based on the Lambertian assumption, and thus, these approaches will suffer from depth ambiguity when addressing the non-Lambertian effect, as demonstrated in Fig. \ref{fig:Result3} (second case). In this paper, we address the problem of large correspondence gap with a non-local attention mechanism to capture large gap correspondences. Since we do not rely on depth information, the proposed method shows higher reconstruction quality on the non-Lambertian cases.}

\textbf{Reconstruction without explicit depth.} These kind of approaches treat light field reconstruction as the approximation of plenoptic function. In the Fourier domain, the sparse sampling in the angular dimension produces overlaps between the original spectrum and its replicas, leading to aliasing effect~\cite{WuEPICNN2018}. Classical approaches~\cite{chai2000plenoptic,zhang2003spectral} consider a reconstruction filter (usually in a wedge shape) to extract the original signal while filtering the aliasing high-frequency components. For instance, Vagharshakyan \textit{et al.}~\cite{Shearlet} utilized an adapted discrete shearlet transform in the Fourier domain to remove the high-frequency spectra that introduce aliasing effects. Shi \textit{et al.}~\cite{LFfourier} performed DSLF reconstruction as an optimization for sparsity in the continuous Fourier domain.

\begin{figure*}
\begin{center}
\includegraphics[width=1\linewidth]{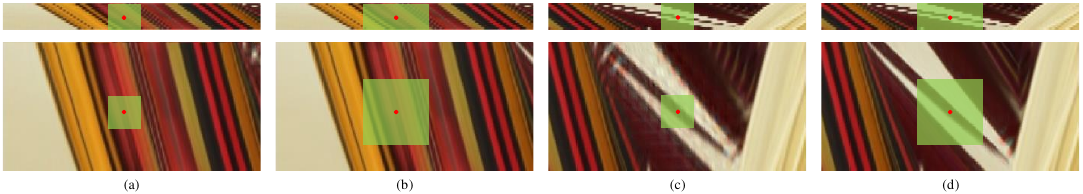}
\end{center}
\vspace{-4mm}
   \caption{Analysis of reconstruction quality in terms of the network receptive field and disparity range of the scene. For a scene with small disparities, both networks (a) with small receptive field ($27\times27$ pixels) and (b) with large receptive field ($53\times53$ pixels) are able to reconstruct high-quality light field. However, for a scene with large disparities, network with small receptive field suffers from severe aliasing effects, as shown in (c). While network with large receptive field can still produce plausible results, as shown in (d). We show the sparely-sampled inputs on the top row and the reconstructed on the bottom. The receptive field of each network is visualized with green box. The input EPIs are stretched along the angular dimension for better demonstration.}
\label{fig:RF}
\vspace{-2mm}
\end{figure*}

Inspired by the success of deep learning in computer vision, depth-independent plenoptic reconstruction approaches~\cite{LFCNN,WuEPICNN2018} were widely investigated by using deep convolution networks. Specifically, Zhu \emph{et al}.~\cite{zhu2019revisiting} proposed an auto-encoder that combines convolutional layers and convLSTM layers~\cite{shi2015convolutional}. For explicitly addressing the aliasing effects, Wu \emph{et al}.~\cite{WuEPICNN2018} took advantage of the clear texture structure of the EPI and proposed a ``blur-restoration-deblur'' framework. However, when applying a large blur kernel for large disparities, this approach fails to recover the high-frequency details, and thus leading to blur effect. \wgc{Liu \textit{et al.}~\cite{liu2020multi} further applied a multi-stream network that takes 3D EPIs in different directions as input. In addition to extracting slices from the plenoptic function,} Yeung \textit{et al.}~\cite{YeungECCV2018} directly fed the entire 4D light field into a pseudo 4D convolutional network, and proposed a \wgc{novel spatial-angular alternating convolution} to iteratively refine the angular dimensions of the light field. \wgc{Jin \textit{et al.}~\cite{jin2020light} further extended the spatial-angular alternating convolution to the problem of compressive light field reconstruction. Wang \textit{et al.}~\cite{wang2020high} applied pseudo 4D convolution to reconstruct the two angular dimensions of the input light field sequentially.} Wu \emph{et al}.~\cite{wu2019learning} proposed an evaluation network for EPIs with different shear amount, termed as sheared EPI structure, \wgc{and further boosted its performance with an end-to-end optimization framework~\cite{wu2021revisiting}.} With this structure, the networks implicitly use depth information to select a well reconstructed EPI. However, the performances of these networks are limited by the finite perceptive field of the convolutional neurons, especially when handling the large disparity problem.

\subsection{Attention Mechanism}
Attention was first built to imitate the mechanism of human perception that mainly focuses on the salient part~\cite{itti1998model,rensink2000dynamic,corbetta2002control}. Vaswani \textit{et al.}~\cite{vaswani2017attention} indicated that the attention mechanism is able to solve the long term dependency problem even without using convolution operation or recurrent neural cell. \wgc{Attention mechanism is typically embedded within a conventional network backbone, such as a VGG-net~\cite{simonyan2015very}, a ResNet~\cite{he2016deep,woo2018cbam} or even a simple multi-layer perceptron (MLP)~\cite{jiang2021transgan}. It encourages the network to focus on the salient parts by assigning an adaptive weight (attention map) to the extracted features.}

Hu \textit{et al.}~\cite{hu2019senet} and Woo \textit{et al.}~\cite{woo2018cbam} proposed to use a global pooling (max-pooling or average-pooling) followed by an MLP to aggregate the entire information in the spatial dimension. \wgc{This attention mechanism enables the network to focus on certain channels in the feature maps. However, the global pooling operation will decimate the high-frequencies in the feature maps~\cite{qin2020fcanet}, which could be unacceptable, especially for a reconstruction task. Alternatively,} Vaswani \textit{et al.}~\cite{vaswani2017attention} proposed to use a weighted average of the responses from all the positions with respect to a certain position for Natural Language Processing (NLP), which is called self-attention. \wgc{Wang \textit{et al.}~\cite{wang2018non-local} further bridged the self-attention for NLP to more general tasks in computer vision, such as video classification. More concretely, a high-dimensional feature map, e.g., a 3D tensor with spatial-temporal dimensions, is reshaped into its 2D form for the operation of matrix multiplication. Different from the convolution, the self-attention allows the element in the feature map to interact with any elements regardless of their distance (range), and thus, is also termed as non-local attention.}

Rather than using the non-local attention mechanism, Wang \textit{et al.}~\cite{wang2019learning,PAM} proposed a parallax attention module to calculate the correspondence between two stereo images along the epipolar line. Tsai \textit{et al.}~\cite{tsaiattention} introduced an attention module in the angular dimension to weight the contribution of each view in a light field. \wgc{Guo \textit{et al.}~\cite{guo2020deep} applied a pixel-wise attention map using convolution layers for adaptively blending feature maps from different frequency components in a light field.}

\wgc{Compared with the attention modules in~\cite{wang2019learning,PAM,tsaiattention,guo2020deep}, the major difference in this paper is that the proposed attention is calculated non-locally in the 2D epipolar plane for each pixel, enabling the network to capture spatial-angular correspondences. That also makes us different from existing non-local attention mechanisms~\cite{wang2018non-local,zhang2018self,vaswani2017attention}, which perform non-local attention across the entire data dimensions. Another significant difference is that we embed a channel-to-angular pixel shuffling operation into the non-local attention mechanism to explicitly achieve the light field reconstruction task.} To the best of our knowledge, our method is the first to apply non-local attention to the light field reconstruction task.

\section{Problem Analysis and Motivation}\label{Sec:Problem}
In this section, we empirically show that the performance of a learning-based light field reconstruction method is closely related to the perception range of its neurons (or convolutional filters) on the epipolar plane, especially when handling a light field with large disparity problem.

Deep neural network is proved to be a \wgc{powerful} technique in solving ill-posed inverse problems~\cite{jin2017deep}. For the light field reconstruction task, both network structure and the disparity range of the scene are significant to the performance of light field reconstruction. Since the disparity range of the scene is unalterable once the light field is acquired, most deep learning-based light field reconstruction methods~\cite{LFCNN,DoubleCNN,WuEPICNN2018,YeungECCV2018} pursue a more appropriate architecture for better performance. Specifically, the depth-based view synthesis methods convert the feature maps into a physically meaningful depth map, while the depth-independent methods directly convert feature maps to novel views. Essentially, both two kinds of approaches adopt convolution operation to generate responses (feature maps) among corresponding pixels.

To quantitatively measure the capability of correspondence capturing, we apply the concept of receptive field introduced in~\cite{Long2014Do,zhou2015object}. The receptive field measures the number of pixels that are connected to a particular filter in the CNN, i.e., the number of correspondence pixels perceived by a certain convolutional filter. 

We analyse the reconstruction qualities of two networks with the same structure (U-net) and the same number of parameters (around 120K) but different receptive fields, as illustrated in Fig. \ref{fig:RF}. For a scene with small disparity (about 3 pixels in the demonstrated example), networks with either small receptive field ($27\times27$ pixels) or large receptive field ($53\times53$ pixels) can reconstruct high angular resolution light fields (EPIs) with view consistency, as shown in Fig. \ref{fig:RF}(a) and Fig. \ref{fig:RF}(b). However, for a scene with large disparity (about 9 pixels), the network with small receptive field is not able to collect enough information from corresponding pixels on the epipolar plane, as shown clearly at the top of Fig. \ref{fig:RF}(c). Since the actual size of the receptive field can be smaller than its theoretical size~\cite{zhou2015object}, the actual receptive field might not be able to cover the disparity range of the input light field, leading to severe aliasing effects in the reconstructed result, as shown at the bottom of Fig. \ref{fig:RF}(c). In contrast, the network with a large receptive field can produce high quality result (Fig. \ref{fig:RF}(d)).

\begin{figure*}
	\begin{center}
		\includegraphics[width=1\linewidth]{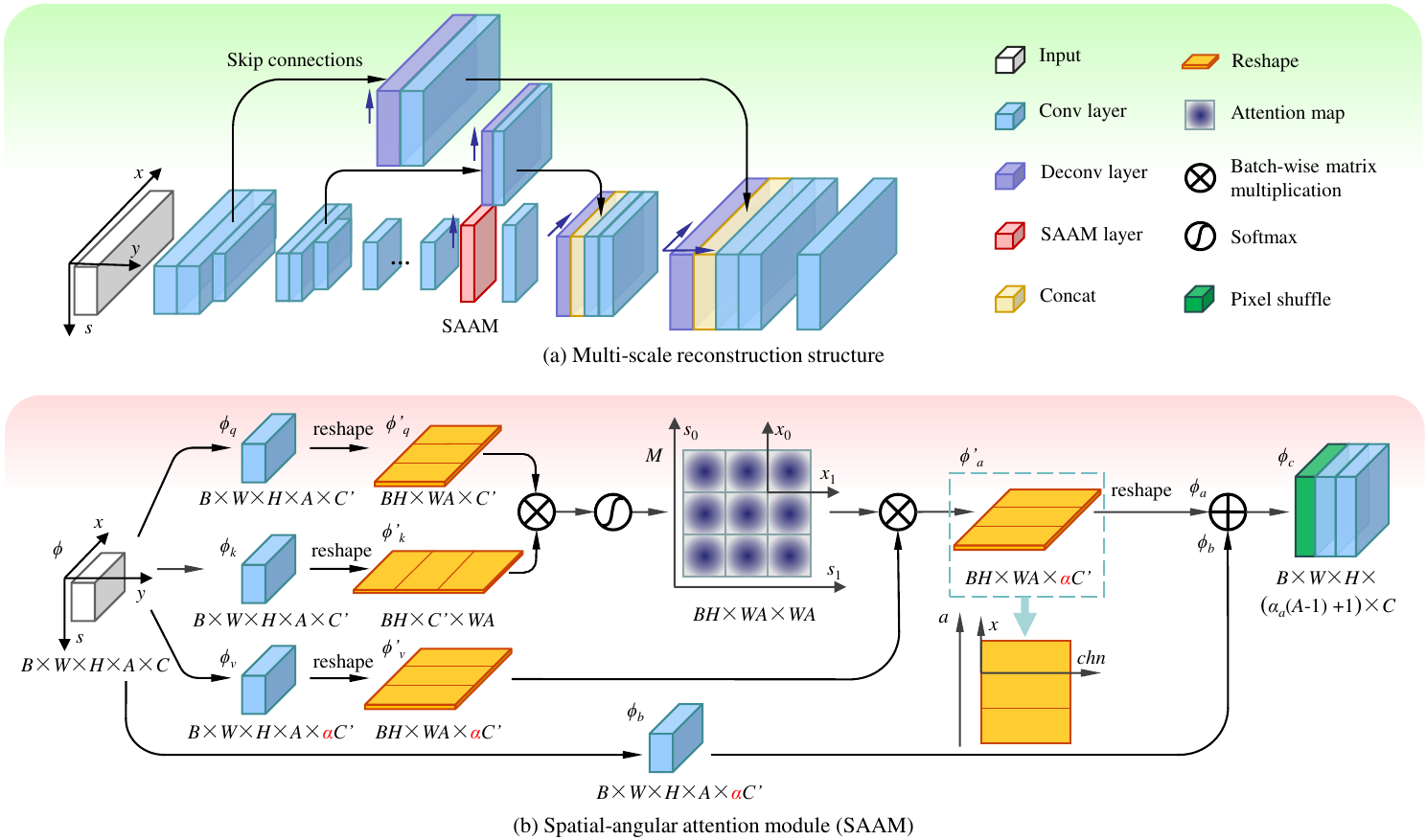}
	\end{center}
	\vspace{-4mm}
	\caption{The proposed Spatial-Angular Attention Network (SAA-Net) is composed of two parts: (a) a Multi-Scale Reconstruction (MSR) structure that maintains the view consistency in the low spatial resolution feature space while preserving fine details in the high spatial resolution feature spaces (Sec. \ref{Sec:Network_archi}); and (b) a Spatial-Angular Attention Module (SAAM) that perceives correspondences on the epipolar plane in a non-local fashion (Sec. \ref{Sec:SAAM}). The input is a 3D slice ($L(u,v,s)$ or $L(v,u,t)$) of the light field.The batch and channel dimensions are omitted in the figure.}
	\label{fig:CNN}
\end{figure*}

Due to the limitation of parameter amount, it is intractable to expand the receptive field by \wgc{pursuing a deeper network or a larger filter size}. The fundamental idea of our proposed light field reconstruction method is to capture the correspondence non-locally across the spatial and angular dimensions of the light field. We achieve this with two features: 1) a spatial-angular attention module that captures non-local correspondence between any two pixels on the epipolar plane; and 2) an encoder-decoder network that reduces the redundancies in the light field to efficiently achieve the non-local perception.

\section{Spatial-Angular Attention Network}\label{Sec:Network}
In this section, we first introduce the overall architecture of the proposed SAA-Net for light field reconstruction, and then introduce the proposed spatial-angular attention module in details. The input of the SAA-Net is a 3D light field slice with two spatial dimensions and one angular dimension, i.e., $L(x,y,s)$ or $L(y,x,t)$. By splitting light fields into 3D slices, the proposed network can be applied to both 3D light fields from a single-degree-of-freedom gantry system and 4D light fields from plenoptic camera and camera array system.

For a 4D light field $L(x,y,s,t)$, we adopt a hierarchical reconstruction strategy similar with that in~\cite{WuEPICNN2018}. The strategy first reconstruct 3D light fields using slices $L_{t^*}(x,y,s)$ and $L_{s^*}(y,x,t)$, then use the synthesized 3D light fields to reconstruct the final 4D light field.

\subsection{Network Architecture}\label{Sec:Network_archi}
We propose a network with Multi-Scale Reconstruction (MSR) structure to maintain view consistency (i.e., continuity in the angular dimension) in the low spatial resolution feature space while preserving fine details in the high spatial resolution feature space. As shown in Fig. \ref{fig:CNN}(a), the backbone of the proposed SAA-Net follows the encoder-decoder structure with skip connections (also known as U-net). But the proposed SAA-Net has two particular differences: 1) In each skip connection, we use a deconvolution layer along the angular dimension before feeding the feature maps to the decoder part; 2) In the encoder part, we use strided convolution with stride only in the spatial dimensions of the light field. Table~\ref{table:Archi} provides the detailed configuration of the proposed SAA-Net.

The \textbf{encoder} part of the SAA-Net generates multi-scale light field features and reduces the redundant information in the spatial dimension to save the computational and GPU memory costs for the non-local perception. We use two convolutional layers (3D) with stride $[2, 2]$ and $[2, 1]$ to downsample the spatial resolution of the light field feature with ratio 4 and 2 along the width and height dimension, respectively. Before each downsampling, two 3D convolutional layers with filter sizes $3\times1\times3$ and $1\times3\times3$ (width, height and angular) are employed to take place of a single convolutional layer with filter size $3\times3\times3$, reducing $1/3$ parameters without performance degradation~\cite{YeungECCV2018}.

The \textbf{skip connections} copy the feature layers before each downsampling layer in the encoder, as shown in Fig. \ref{fig:CNN}(a). For each skip connection, a deconvolution layer (also known as transposed convolution layer) is applied to upsample the feature map in the angular dimension, followed by a $1\times1\times1$ convolution. Since the angular information mainly concentrates on the 2D EPI $E(x,s)$ for reconstructing a 3D light field $L(x,y,s)$, the filter size in each deconvolution layer in the skip connection is set to $3\times1\times7$.

The \textbf{decoder} part of the SAA-Net upsamples the feature map from the spatial-angular attention module (Sec. \ref{Sec:SAAM}) by using two deconvolution layer with stride $[2, 1]$ and $[2, 2]$ in the spatial dimensions (width and height), respectively. The decoder also receives information from the skip connections by concatenating the features from corresponding levels along the channel dimension~\cite{Eilertsen2017HDR}, as shown in Fig. \ref{fig:CNN}(a). We then use two 3D convolutional layers with filter sizes $3\times1\times3$ and $1\times3\times3$ to compress the channel numbers in each level of the decoder. This can be considered as the blending of the light field features from different reconstruction scale. Note that all the reconstructions (upsampling operations) in the angular dimension are implemented in the skip connections and the spatial-angular attention module, where the latter will be introduced in the following subsection.

\subsection{Spatial-Angular Attention Module}\label{Sec:SAAM}
Inspired by the non-local attention mechanism in \cite{wang2018non-local,zhang2018self}, we propose a Spatial-Angular Attention Module (SAAM) \wgc{to disentangle} the disparity information in light field. The main differences between the proposed SAAM and the previous non-local attention~\cite{wang2018non-local,zhang2018self} are as follows: 1) Since the disparity information is encoded in the EPI, the non-local attention mechanism is performed in the 2D epipolar plane rather than the entire 3D space; 2) We model the light field reconstruction with pixel shuffling~\cite{shi2016real} that disentangles the reconstructed angular information from the channel dimension.

A straightforward choice to perform spatial-angular attention is to embed the attention module in each resolution scale of the U-net. However, implementing non-local perception in the full resolution light field (feature map) is intractable due to the high computation complexity and GPU memory cost. Alternatively, we insert the proposed SAAM between the encoder and decoder, i.e., to perform non-local operation in the low-resolution feature space, as shown in Fig. \ref{fig:CNN}. 

Since features in a 3D CNN will be a 5D tensor $\phi\in\mathbb{R}^{B\times W\times H\times A\times C}$ (i.e., batch, width, hight, angular and channel), we first apply two convolution layers with kernel size $1\times1\times1$ to produce two feature layers $\phi_q$ and $\phi_k$ with size of $B\times W\times H\times A\times C'$. The channel number $C'$ is set to be $\frac{C}{8}$ (i.e., $C'=6$ in our implementation) for computation efficiency. Then the feature layers $\phi_q$ and $\phi_k$ are reshaped into 3D tensors $\phi'_q$ and $\phi'_k$ of size $BH\times WA\times C'$ and $BH\times C'\times WA$, respectively. In this way, we merge the angular and width dimensions ($s$ and $x$ or $t$ and $y$ in a light field) together to implement the non-local perception on the epipolar plane. 

We apply batch-wise matrix multiplication between $\phi'_q$ and $\phi'_k$ and use a softmax function to produce \wgc{an attention map} $M$ as illustrated in Fig. \ref{fig:CNN}(b). The attention map is composed of $BH$ matrices with shape $WA\times WA$. Each matrix can be considered as a 2D expansion map of a 4D tensor $M'\in\mathbb{R}^{W\times A\times W\times A}$ (the batch and height dimensions are neglected). The point $M'(x_0,s_0,x_1,s_1)$ indicates the response of light field position $L(x_0,y,s_0)$ to position $L(x_1,y,s_1)$ in the latent space. In other words, the attention map is able to capture correspondence among all the views in the input 3D light field. 

\begin{figure}
\begin{center}
\includegraphics[width=1\linewidth]{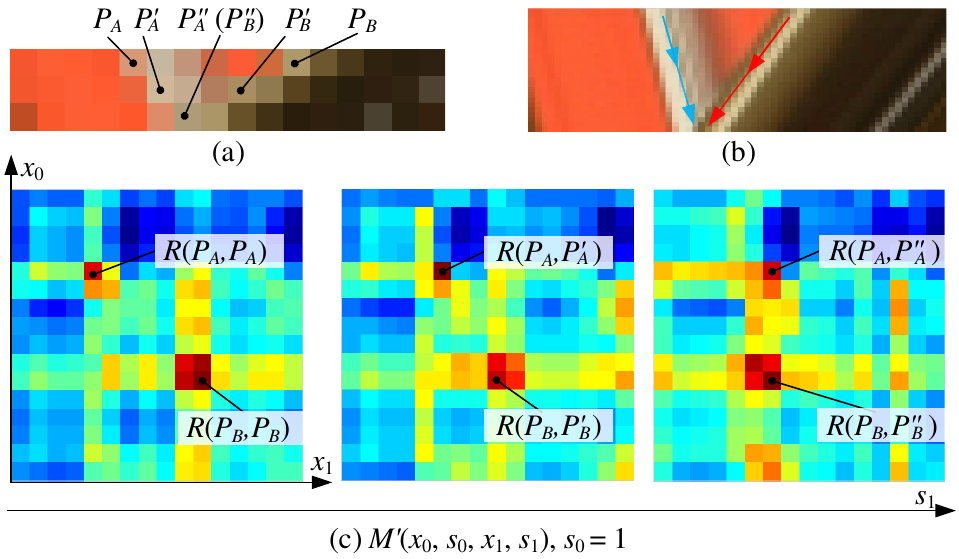}
\end{center}
\vspace{-4mm}
   \caption{Visualization of the attention map before the softmax function. (a) An EPI with a foreground point $P_A$ and a background point $P_B$; (b) The corresponding high spatial-angular resolution EPI; (c) Three sub-maps extracted from the attention map. A point will have a strong response at the location of its correspondence in the attention map.}
\label{fig:att_map}
\end{figure}

\begin{table}
\caption{Detail configuration of the proposed SAA-Net, where $k$ denotes the kernel size, $s$ the stride, $chn$ the number of channels, Conv the 3D convolution layer, Deconv the 3D deconvolution layer and Concat the concatenation.}
\label{table:Archi}
\begin{center}
\begin{tabular}{l|cccc}
Layer & $k$ & $s$ & $chn$ & Input\\
\hline
\multicolumn{5}{c}{Encoder}\\
\hline
Conv1$\_1$ & $3\times1\times3$ & $[1,1,1]$ & 1/24 & $L(x,y,s)$\\
Conv1$\_2$ & $1\times3\times3$ & $[1,1,1]$ & 24/24 & Conv1$\_1$\\
Conv1$\_3$ & $3\times3\times1$ & $[2,2,1]$ & 24/48 & Conv1$\_2$\\
Conv2$\_1$ & $3\times1\times3$ &$[1,1,1]$ & 48/48 & Conv1$\_3$\\
Conv2$\_2$ & $1\times3\times3$ & $[1,1,1]$ & 48/48 & Conv2$\_1$\\
Conv2$\_3$ & $3\times1\times1$ & $[2,1,1]$ & 48/96 & Conv2$\_2$\\
Conv3$\_1$ & $1\times1\times1$ & $[1,1,1]$ & 96/48 & Conv2$\_3$\\
Conv3$\_2$ & $3\times1\times3$ & $[1,1,1]$ & 48/48 & Conv3$\_1$\\
Conv3$\_3$ & $1\times3\times3$ & $[1,1,1]$ & 48/48 & Conv3$\_2$\\
Conv3$\_4$ & $3\times1\times3$ & $[1,1,1]$ & 48/48 & Conv3$\_3$\\
Conv3$\_5$ & $1\times3\times3$ & $[1,1,1]$ & 48/48 & Conv3$\_4$\\
\hline
\multicolumn{5}{c}{Skip connection}\\
\hline
Deconv4$\_1$ & $3\times1\times7$ & $[1,1,\alpha]$ & 24/24 & Conv1$\_2$\\
Conv4$\_2$ & $1\times1\times1$ & $[1,1,1]$ & 24/24 & Deconv4$\_1$\\
Deconv5$\_1$ & $3\times1\times7$ & $[1,1,\alpha]$ & 48/48 & Conv2$\_2$\\
Conv5$\_2$ & $1\times1\times1$ & $[1,1,1]$ & 48/48 & Deconv5$\_1$\\
\hline
\multicolumn{5}{c}{SAAM}\\
\hline
\multicolumn{5}{c}{Decoder}\\
\hline
Conv6$\_1$ & $1\times1\times1$ & $[1,1,1]$ & 48/96 & SAAM\\
Deconv6$\_2$ & $4\times1\times1$ & $[2,1,1]$ & 96/48 & Conv6$\_1$\\
Concat1 & - & - & -  & Conv6$\_1$; Conv4$\_2$\\
Conv6$\_3$ & $3\times1\times3$ & $[1,1,1]$ & 48/48 & Concat1\\
Conv6$\_4$ & $1\times3\times3$ & $[1,1,1]$ & 48/48 & Conv6$\_3$\\
Deconv7$\_1$ & $4\times4\times1$ & $[2,2,1]$ & 48/24 & Conv6$\_4$\\
Concat2 & - & - & -  & Conv7$\_1$; Conv5$\_2$\\
Conv7$\_2$ & $3\times1\times3$ & $[1,1,1]$ & 24/24 & Concat2\\
Conv7$\_3$ & $1\times3\times3$ & $[1,1,1]$ & 24/24 & Conv7$\_2$\\
Conv8 & $3\times3\times3$ & $[1,1,1]$ & 24/1 & Conv7$\_3$\\
\end{tabular}
\end{center}
\end{table}

We demonstrate the non-local perception of the proposed SAAM by visualizing a part of the attention map before the softmax function as shown in Fig. \ref{fig:att_map}. In this example, there are two points $P_A$ and $P_B$ with remarkable visual features as shown in Fig. \ref{fig:att_map}(a), and their corresponding points in other views are marked as $P_A'$ ($P_A''$) and $P_B'$ ($P_B''$). As the viewpoint changes along the angular dimension, the background point $P_A$ will be occluded by the foreground point $P_B$, which is demonstrated more obviously in Fig. \ref{fig:att_map}(b). Fig. \ref{fig:att_map}(c) shows three sub-maps extracted from the attention map $M'$ with $s_0=1$ and $s_1=1,2,3$, respectively. It can be clearly seen that a point will have the highest response at the location of its correspondence in the attention map. For instance, the response $R(P_B,P_B')$ at the location $M'(11,1,9,2)$ for the corresponding patch $(P_B,P_B')$ (the middle sub-figure of Fig. \ref{fig:att_map}(c)), and the response $R(P_B,P_B'')$ at the location $M'(11,1,7,3)$ for the corresponding patch $(P_B,P_B'')$ (the right sub-figure of Fig. \ref{fig:att_map}(c)). For the occluded point $P_A$, the location of the maximum response changes from $M'(5,1,5,1)$ to $M'(5,1,7,3)$. In this case, the attention module is able to locate the occluded point $P_A''$ through its surrounding pixels. More demonstrations of spatial-angular attention map can be found in Sec. \ref{Sec:attention_map}.

Feature $\phi_v$ and $\phi_b$ are obtained by another two $1\times1\times1$ convolutions in a similar manner as that for $\phi_q$ and $\phi_k$. The main difference is that the channel numbers of these two feature layers are $\alpha C'$, where, $\alpha$ denotes the reconstruction factor (upsampling scale in the angular dimension) of the network. Another batch-wise matrix multiplication is applied between the attention map $M$ and $\phi'_v$ (reshaped from $\phi_v$), resulting a 3D tensor $\phi'_a\in\mathbb{R}^{BH\times WA\times\alpha C'}$. We then reshape $\phi'_a$ into a 5D tensor $\phi_a\in\mathbb{R}^{B\times W\times H\times A\times\alpha C'}$.

\begin{figure}
\begin{center}
\includegraphics[width=.85\linewidth]{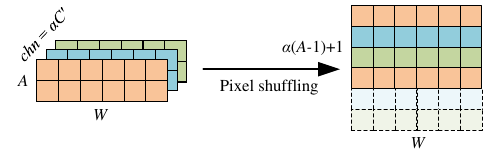}
\end{center}
\vspace{-4mm}
\caption{\wgc{We employ channel-to-angular pixel shuffling for light field reconstruction. To avoid view extrapolation effect in the reconstructed light filed, we remove the last $\alpha-1$ elements in the angular dimension.}}
\label{fig:pixelshuffle}
\end{figure}

\begin{table}
\caption{Detail configuration of the proposed Spatial-Angular Attention Module (SAAM), where MatMul denotes the matrix multiplication and Add the element-wise addition.}
\label{table:Archi_SAAM}
\begin{center}
\begin{tabular}{l|ccc}
Layer & $k$ & $chn$ & Input\\
\hline
Conv1 & $1\times1\times1$ & $C$/$C'$ & Encoder\\
Conv2 & $1\times1\times1$ & $C$/$C'$ & Encoder\\
Conv3 & $1\times1\times1$ & $C$/$\alpha C'$ & Encoder\\
Conv4 & $1\times1\times1$ & $C$/$\alpha C'$ & Encoder\\
Reshape1 &- & $C'$/- & Conv1 \\
Reshape2 &- & $C'$/- & Conv2 \\
Reshape3 &- & $C'$/- & Conv3 \\
MatMul1 & - & - & Reshape1; Reshape2 \\
Softmax & - & - & MatMul1\\
MatMul2 & - & - & Softmax; Reshape3 \\
Reshape4 &- & -/$\alpha C'$ & MatMul2 \\
Add & - & $\alpha C'$/$\alpha C'$  & Reshape4; Conv4\\
Pixel shuffle & $3\times1\times7$  & $\alpha C'$/$C'$ & Add\\
Conv5 & $7\times1\times1$  & \wgc{$C'$/$C$} & Pixel shuffle \\
Conv6 & $1\times1\times7$  & $C$/$C$ & Conv5\\
\end{tabular}
\end{center}
\end{table}

Using the aforementioned SAAM, the reconstructed angular information can be well encoded in the channel dimension of the 5D tensor. By disentangling it with channel-to-angular pixel shuffling, we can reconstruct a high angular resolution light field (feature map) in a non-local manner. Specifically, we first multiply the feature layer $\phi_a$ by a trainable scale parameter (initialized as 0) and add back to the feature layer $\phi_b$. We then apply the channel-to-angular pixel shuffling and reconstruct a 5D tensor $\phi_c\in\mathbb{R}^{B\times W\times H\times (\alpha (A-1)+1)\times C'}$. \wgc{As illustrated in Fig. \ref{fig:pixelshuffle}, the channel-to-angular pixel shuffling rearranges elements from channel dimension to angular dimension}\footnote{Different from typical spatial super-resolution, upsampling a light field by $\alpha\times$ generates a light field of $\alpha (A-1)+1$ views. So we simply remove the last $\alpha-1$ views from the reconstructed light field.}. The final output of the SAAM is generated by two convolutional layers with kernel sizes of $7\times1\times1$ and $1\times1\times7$, respectively. 

By combining the proposed SAAM with the feature maps in the skip connections, the network is able to reconstruct light field with view consistency while also preserving the high frequency components. Detailed parameter setting of the SAAM is listed in Table \ref{table:Archi_SAAM}.


\section{Network Training}\label{Sec:training}
\subsection{Spatial-Angular Perceptual Loss}
Typical learning-based light field reconstruction or view synthesis methods optimize the network parameters by formulating a pixel-wise loss between the inferred image and the desired view (or EPI~\cite{WuEPICNN2018}). Recently, researches~\cite{zhou2018stereo,zhang2019image,meng2019high,mildenhall2019local} show that formulating the loss function in the high-level feature space will motivate the restoration of high-frequency components. This high-level feature loss, also known as perceptual loss, can be computed from part of the feature layers in the autologous network~\cite{dosovitskiy2016generating} or other pre-trained networks~\cite{Johnson2016Perceptual}, such as the commonly-used VGG network~\cite{simonyan2015very}.

In this paper, we propose a spatial-angular perceptual loss that is specifically designed for the high-dimensional light field data. Existing approaches~\wgc{\cite{mildenhall2019local,meng2019high,meng2021light}} for light field reconstruction apply perceptual loss between 2D sub-aperture images, neglecting the view consistency constraint in the angular dimension. Alternatively, we propose to use a 3D light field encoder to map the 3D light fields into high-dimensional feature tensors (width, height, angular and channel). 

\wgc{To achieve this, we design another 3D encoder-decoder network (auto-encoder) that learns to extract the high-level features for the proposed spatial-angular perceptual loss}\footnote{The architecture of the 3D auto-encoder for the perceptual loss is different with that of the SAA-Net.}. Note that the auto-encoder can be also generalized to a 4D form. But given that some light field datasets have only one angular dimension (e.g., light fields from gantry system in~\cite{ICME2018}) and the proposed SAA-Net also takes 3D light field as its input, we only adopts 3D convolution in the encoder and decoder.

\begin{figure}
\begin{center}
\includegraphics[width=1\linewidth]{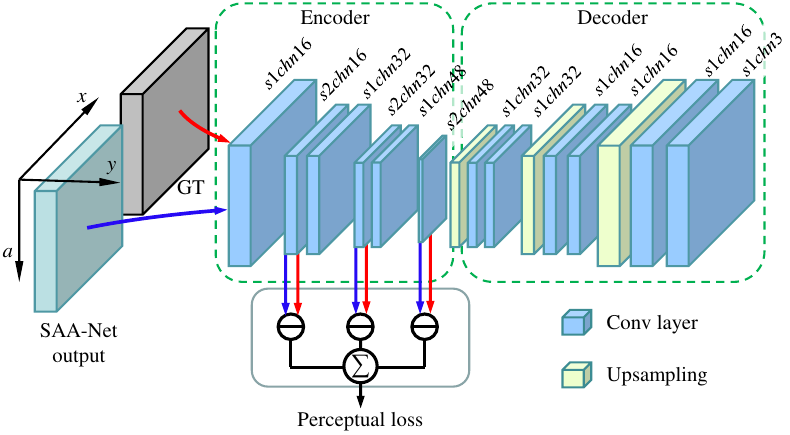}
\end{center}
\vspace{-4mm}
   \caption{Architecture of the 3D encoder-decoder network designed for the proposed spatial-angular perceptual loss.}
\label{fig:AE}
\end{figure}

\wgc{The proposed auto-encoder for the spatial-angular perceptual loss uses the output of the SAA-Net or the desired 3D light field (ground truth) as input, as shown in Fig. \ref{fig:AE}. It has 12 convolutional layers ($\phi_{ae}^{(l)}, l=1,2,\cdots,12$) with kernel size $3\times3\times3$ and 3 bilinear upsampling layers. The encoder part includes the first 6 convolutional layers, where 3 convolutional layers ($\phi_{ae}^{(2)}, \phi_{ae}^{(4)}, \phi_{ae}^{(6)}$) with stride 2 in each dimension are used to compress the light field from low-level pixel space into high-level feature space. The decoder part includes the last 6 convolutional layers and 3 upsampling layers to restore the light field from the latent representations. Detailed configuration of each layer (stride and number of channels) is shown in Fig. \ref{fig:AE}.}

\wgc{The auto-encoder learns how to extract high-level features through unsupervised learning, i.e., the network learns to predict the input 3D light field. The objective for training the auto-encoder is $\mathcal{L}_{AE}(L_{HR})=\Vert f_{AE}(L_{HR})-L_{HR} \Vert_1$, where $f_{AE}$ denotes the auto-encoder. We use the same training dataset (Sec. \ref{Sec:train_data}), learning rate and optimizer (Sec. \ref{Sec:train_details}) as those for the SAA-Net.}

\wgc{To compute the final spatial-angular perceptual loss, we feed the output of the SAA-Net $\hat{L}_{HR}$ as well as the ground truth light field $L_{HR}$ to the auto-encoder, as shown in Fig. \ref{fig:AE}. And the spatial-angular perceptual loss is defined as follows}
$$
\mathcal{L}_{feat}(\hat{L}_{HR},L_{HR})=\sum\limits_{l=2,4,6}\lambda_{feat}^{(l)}\Vert\phi_{ae}^{(l)}(\hat{L}_{HR})-\phi_{ae}^{(l)}(L_{HR})\Vert_1,
$$
where $\phi_{ae}^{(l)}(\cdot)$ $(l={2,4,6})$ denotes the \wgc{feature maps obtained from the $l$th layer in the encoder}, and $\lambda_{feat}=0.2, 0.2, 0.1$ is a set of hyperparameters for the proposed spatial-angular perceptual loss.

To prevent the potential possibility that different light field patches are mapped to the same feature vector~\cite{dosovitskiy2016generating}, our loss function also contains a pixel-wise term $\mathcal{L}_{pix}$ using Mean Absolute Error (MAE) between $\hat{L}_{HR}$ and $L_{HR}$, i.e.,
$$\mathcal{L}_{pix}(\hat{L}_{HR},L_{HR})=\Vert \hat{L}_{HR}-L_{HR} \Vert_1.$$
Then the final loss function $\mathcal{L}_{SAA}$ for training the SAA-Net is defined as
$$
\mathcal{L}_{SAA}=\mathcal{L}_{pix}+\mathcal{L}_{feat}.
$$
The two terms are weighted by the set of hyperparameters $\lambda_{feat}$ in the perceptual loss.

\subsection{Training Data}\label{Sec:train_data}
\begin{figure}
\begin{center}
\includegraphics[width=1\linewidth]{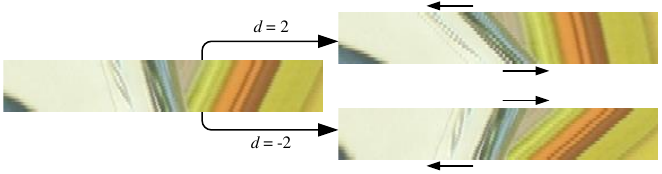}
\end{center}
\vspace{-4mm}
   \caption{An illustration of training data augmentation using shearing operation. For clear display, one of the spatial dimension in the 3D light field is ignored.}
\label{fig:data_aug}
\vspace{-1mm}
\end{figure}

We use light fields from the Stanford (New) Light Field Archive~\cite{StanfordLFdatasets} as the training dataset, which contains 12 light fields\footnote{The light field \textit{Lego Gantry Self Portrait} is excluded from the training dataset since the moving camera may influence the reconstruction performance.} with $17\times17$ views. Since the network input is 3D light fields, we can extract 17 $L(x,y,s)$ and 17 $L(y,x,t)$ in each 4D light field set. Similar with the data augmentation strategy proposed in~\cite{zhu2019revisiting}, we augment the extracted 3D light fields using shearing operation~\cite{Ng2005Fourier}
$$L_d(x,y,s)=L(x+(s-\frac{S}{2})\cdot d,y,s),$$
where $S$ is the angular resolution of the 3D light field $L(x,y,s)$, and $L_d(x,y,s)$ is the resulting 3D light field with shear amount $d$. $L_d(y,x,t)$ can be obtained following a similar manner. In practice, we use two shear amounts $d=\pm2$. The shearing-based data augmentation increases the number of training examples by 2 times. More importantly, the disparity effects in the augmented light field will be more obvious as shown in Fig. \ref{fig:data_aug}, enabling the network to address the large disparity problem.

To accelerate the training procedure, the extracted 3D light fields are cropped into sub-light fields with a spatial resolution of $64\times24$ (width and height for $L(x,y,s)$ or height and width for $L(y,x,t)$) and a stride of 40 pixels. About $6.7\times10^5$ examples can be extracted from the 3D light fields (original and sheared).

\vspace{-1mm}
\subsection{Implementation Details}\label{Sec:train_details}
We train two models with reconstruction factors $\alpha=3,4$. The input/output angular resolution of the training samples for these two models are 6/16 and 5/17, respectively. Although the reconstruction factor of the network is fixed, we can achieve a flexible upsampling rate through network cascade. The training is performed on the Y channel (i.e., the luminance channel) of the YCbCr color space. We initialize the weights of both convolution and deconvolution layers by drawing randomly from a Gaussian distribution with a zero mean and a standard deviation of $1\times10^{-3}$, and the biases by zero. The network is optimized by using ADAM solver~\cite{Kingma2014Adam} with learning rate of $1\times10^{-4}$ ($\beta_1=0.9$, $\beta_2=0.999$) and mini-batch size of 28. The training model is implemented in the \emph{Tensorflow} framework~\cite{TensorFlow}. The network converges after $8\times10^5$ steps of backpropagation, taking about 35 hours on a NVIDIA Quadro GV100.

\begin{figure*}
\begin{center}
\includegraphics[width=1\linewidth]{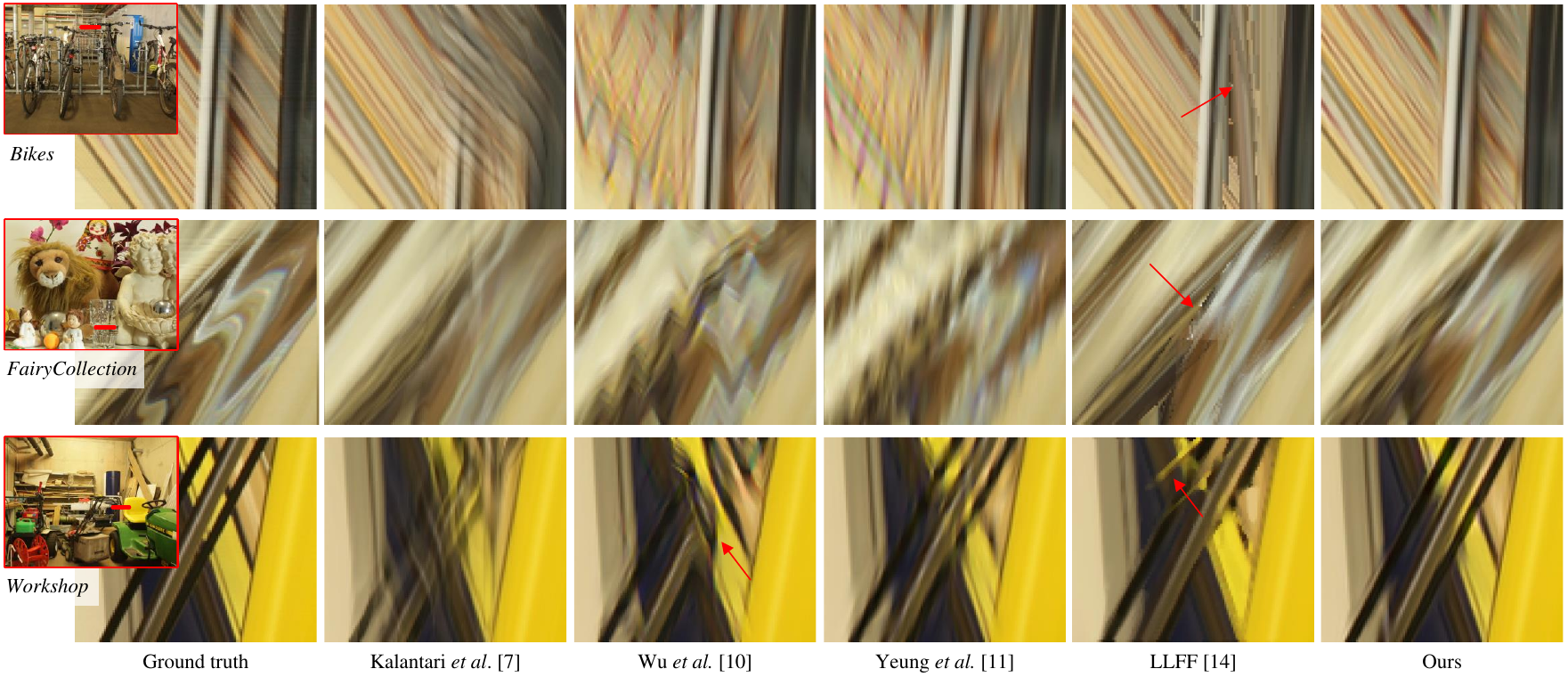}
\end{center}
\vspace{-4mm}
   \caption{Comparison of the results (reconstructed EPIs) on the light fields from the MPI Light Field Archive~\cite{kiran2017towards} ($16\times$ upsampling).}
\vspace{-2mm}
\label{fig:Result3}
\end{figure*}
                                                                                                                                                                                                                                                                                                                                                                                                                                                                           
\section{Evaluations}
In this section, we evaluate the proposed SAA-Net on various kinds of light fields, including those from both gantry systems and from plenoptic camera (Lytro Illum~\cite{Lytro}). \wgc{We mainly compare our method with six state-of-the-arts learning-based methods including Kalantari~\textit{et al.}~\cite{DoubleCNN} (depth-based), LLFF~\cite{mildenhall2019local} (MPI representation), Wu~\textit{et al.}~\cite{WuEPICNN2018}, Yeung~\textit{et al.}~\cite{YeungECCV2018}, HDDRNet~\cite{meng2019high} and DA$^2$N~\cite{wu2021revisiting} (without explicit depth).} To fully demonstrate the effectiveness of our design choices, we also perform ablation studies by training our network without the SAAM, without the MSR structure and without the spatial-angular perceptual loss, respectively. The quantitative evaluations is reported by measuring the average PSNR and SSIM~\cite{SSIM} values over the synthesized views of the luminance channel in the YCbCr space. Please refer to the submitted video for more qualitative results.

\begin{table*}
\caption{Quantitative results (PSNR/SSIM) of reconstructed light fields on the light fields from the MPI Light Field Archive~\cite{kiran2017towards}.}
\label{table:Result3}
\vspace{-3mm}
\begin{center}
\begin{tabular}{p{2.4cm}|c|p{1.9cm}<{\centering} p{1.9cm}<{\centering} p{1.9cm}<{\centering} p{1.9cm}<{\centering} p{1.9cm}<{\centering} p{1.9cm}<{\centering}}
& Scale & \textit{Bikes} & \textit{FairyCollection} & \textit{LivingRoom} & \textit{Mannequin} & \textit{WorkShop} & Average\\
\hline
Kalantari~\textit{et al.}~\cite{DoubleCNN}& \multirow{7}*{$8\times$} & 34.83 / 0.969 & 36.66 / 0.977 & 46.35 / 0.991 & 40.62 / 0.983 & 38.66 / 0.986 & 39.42 / 0.981\\
Wu~\textit{et al.}~\cite{WuEPICNN2018}& & 38.39 / 0.990 & 40.32 / 0.992 & 45.48 / 0.996 & 43.26 / 0.995 & 41.55 / 0.995  & 41.80 / 0.994\\
Yeung~\textit{et al.}~\cite{YeungECCV2018} & & 39.55 / 0.993 & 40.25 / 0.993 & 47.32 / 0.997 & 44.49 / 0.996 & 43.17 / 0.996 & 42.96 / 0.995\\
LLFF~\cite{mildenhall2019local} & & 36.84 / 0.985 & 40.14 / 0.989 & 46.85 / 0.990 & 43.23 / 0.989 & 41.79 / 0.991 & 41.77 / 0.989\\
\wgc{HDDRNet~\cite{meng2019high}} & &   38.97 / 0.994 & 40.20 / 0.994 & 43.65 / 0.997 & 42.21 / 0.996 & 42.01 / 0.997 & 41.41 / 0.996\\
\wgc{DA$^2$N~\cite{wu2021revisiting}}&& 39.14 / 0.992 & 41.00 / 0.993 & 46.30 / 0.996 & 44.17 / 0.996 & 43.09 / 0.996 & 42.74 / 0.994\\
Our proposed & &\textbf{40.35 / 0.994} & \textbf{42.22 / 0.995} & \textbf{48.01 / 0.997} &\textbf{44.99 / 0.996} &\textbf{45.29 / 0.997}  &\textbf{44.17 / 0.996}\\
\hline
Kalantari~\textit{et al.}~\cite{DoubleCNN}& \multirow{12}*{$16\times$} & 30.67 / 0.935 & 32.39 / 0.952 & 41.62 / 0.973 & 37.15 / 0.970  & 33.94 / 0.971 & 35.15 / 0.960\\
Wu~\textit{et al.}~\cite{WuEPICNN2018}& & 31.22 / 0.951 & 30.33 / 0.942 & 42.43 / 0.991 & 39.53 / 0.989 & 33.49 / 0.977 & 35.40 / 0.970\\
Yeung~\textit{et al.}~\cite{YeungECCV2018} & & 32.67 / 0.967 & 31.82 / 0.969 & 43.54 / 0.993 & 40.82 / 0.992 & 37.21 / 0.988 & 37.21 / 0.982\\
LLFF~\cite{mildenhall2019local} & & 34.95 / 0.963 & 34.01 / 0.966 & 44.73 / 0.987 & 39.92 / 0.985 & 37.61 / 0.985 & 38.24 / 0.977\\
\wgc{HDDRNet~\cite{meng2019high}} & & 33.97 / 0.976 & 35.08 / 0.979 & 44.83 / 0.997 & 40.60 / 0.993 & 38.54 / 0.992 & 38.60 / 0.987\\
\wgc{DA$^2$N~\cite{wu2021revisiting}} & & 35.79 / 0.984 & 36.23 / 0.981 & 45.91 / 0.996 & 40.83 / 0.992 & 40.11 / 0.994 & 39.77 / 0.990\\
Our proposed & &\textbf{36.54 / 0.987} & \textbf{37.62 / 0.987} & \textbf{47.25 / 0.997} & \textbf{42.08 / 0.994}  & \textbf{40.55 / 0.994} & \textbf{40.81 / 0.992}\\
\cline{1-1}\cline{3-8}
\wgc{MSR structure-2} & & 34.13 / 0.972 & 35.72 / 0.984 & 46.21 / 0.996 & 40.97 / 0.993 & 38.66 / 0.991 & 39.14 / 0.987\\
\wgc{MSR structure-3} & & 34.20 / 0.973 & 35.53 / 0.984 & 46.79 / 0.996 & 41.04 / 0.993 & 38.85 / 0.991 & 39.28 / 0.987\\
\wgc{MSR structure-4} & & 34.45 / 0.977 & 35.05 / 0.977 & 47.00 / 0.995 & 41.01 / 0.993 & 38.26 / 0.991 & 39.15 / 0.987\\
\wgc{w/o MSR structure}&& 33.76 / 0.975 & 34.35 / 0.976 & 46.07 / 0.995 & 40.19 / 0.991 & 38.05 / 0.992 & 38.48 / 0.986\\
\wgc{w/o SAP loss}    & & 36.43 / 0.987 & 37.29 / 0.986 & 47.10 / 0.997 & 41.81 / 0.993 & 40.18 / 0.993 & 40.56 / 0.991\\
\end{tabular}
\begin{tablenotes}
\textit{``MSR structure-X'' denotes the proposed network that has $X$ series of the MSR structure without using the SAAM.}
\end{tablenotes}
\end{center}
\vspace{-5mm}
\end{table*}

\subsection{Evaluations on Light Fields from Gantry Systems}
In this experiment, the comparisons are performed on light fields from the MPI Light Field Archive~\cite{kiran2017towards} (\wgc{$1\times101$ views of resolution $960\times720$, $1\times97$ views for evaluation}) and the CIVIT Dataset~\cite{ICME2018} ($1\times193$ views of resolution $1280\times720$) with upsampling scales $8\times$ and $16\times$. The performances with respect to both angular sparsity and non-Lambertian are taken into consideration. Since the vanilla version of the network by Yeung~\textit{et al.}~\cite{YeungECCV2018}\footnote{In the modified implementation, every 8 (6) views are applied to reconstruct (synthesize) a 3D light field of 22 (21) views for the networks of reconstruction factor $\alpha=3$ ($\alpha=4$).} and Meng \textit{et al.}~\cite{meng2019high} (HDDRNet) were specifically designed for 4D light fields, we modify their convolutional layers to fit the 3D input while keeping its network architecture unchanged. The networks by Kalantari~\textit{et al.}~\cite{DoubleCNN}, Yeung et al.~\cite{YeungECCV2018}, Mildenhall~\textit{et al.}~\cite{mildenhall2019local} (LLFF) and Meng \textit{et al.}~\cite{meng2019high} (HDDRNet) are re-trained using the same training dataset as our SAA-Net. Due to the particularity of the training datasets in~\cite{wu2021revisiting}, we do not retrain the DA$^2$N. We perform network cascade to achieve different upsampling scales, i.e., two cascades for $8\times$ ($16\times$) upsampling using a network of reconstruction factor $\alpha=3$ ($\alpha=4$).

Fig. \ref{fig:Result3} shows the reconstruction results on three light fields, \textit{Bikes}, \textit{FairyCollection} and \textit{WorkShop}, from the MPI Light Field Archive~\cite{kiran2017towards} with upsampling scale $16\times$ (disparity range up to 33.5px). The first and the third cases have complex occlusion structures, as shown in the top and the bottom row in Fig. \ref{fig:Result3}. The baseline methods~\cite{DoubleCNN,WuEPICNN2018,YeungECCV2018,mildenhall2019local} fail to reconstruct the complex structures. Among them, the depth and learning-based approach~\cite{DoubleCNN} and the MPI-based approach~\cite{mildenhall2019local} fail to estimate proper occlusion relations between the foregrounds and the backgrounds. The second scene is a non-Lambertian case, i.e., a refractive glass before the toys. The approach by Kalantari~\textit{et al.}~\cite{DoubleCNN} cannot reconstruct the refractive object. And the EPIs reconstructed by the baseline methods~\cite{WuEPICNN2018,YeungECCV2018} appear severe aliasing effects due to the limited receptive field. The MPI-based approach LLFF~\cite{mildenhall2019local} is able to reconstruct the non-Lambertian effects in this case, but produces ghosting artifacts as marked by the red arrow in Fig. \ref{fig:Result3}. 

\begin{figure*}
\begin{center}
\includegraphics[width=1\linewidth]{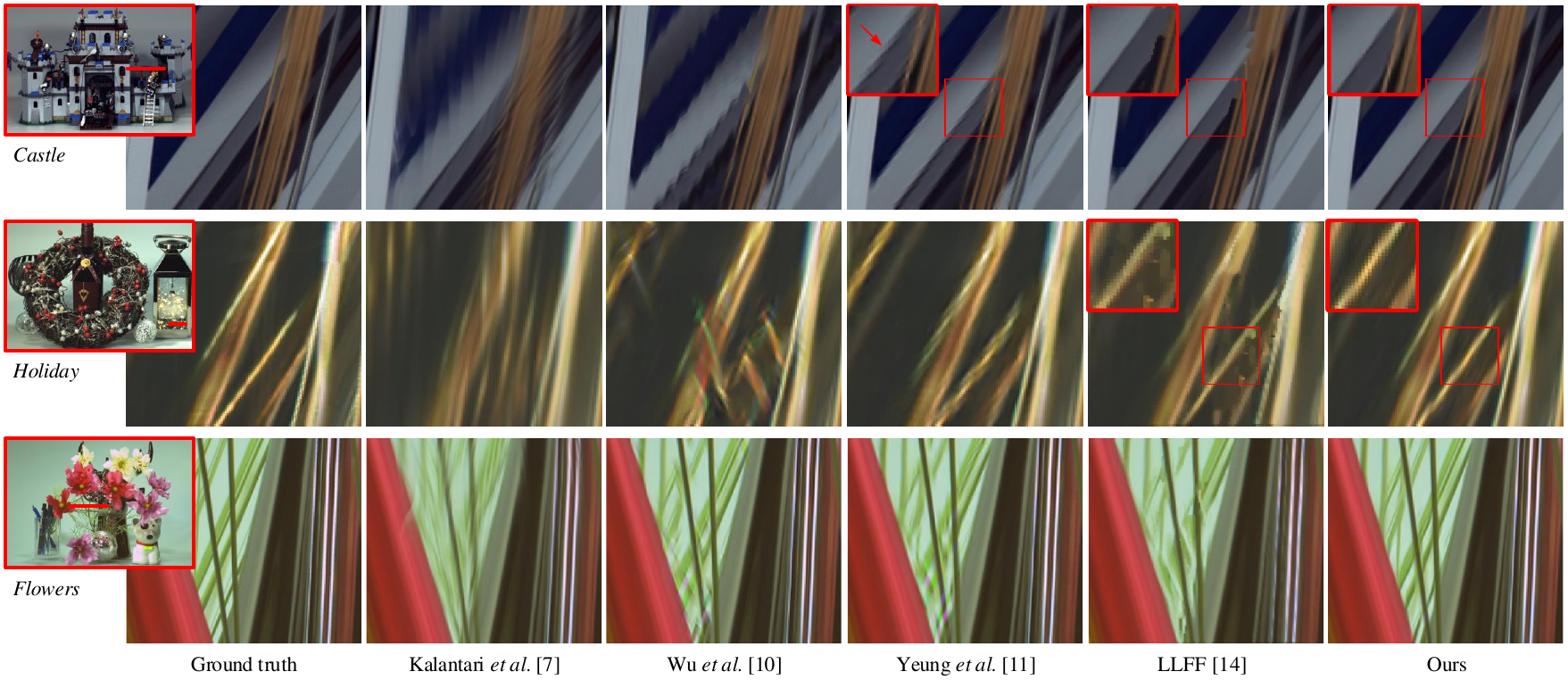}
\end{center}
\vspace{-4mm}
   \caption{Comparison of the results on the light fields from the CIVIT Dataset~\cite{ICME2018} ($16\times$ upsampling).}
\label{fig:Result2}
\vspace{-3mm}
\end{figure*}

Fig. \ref{fig:Result2} shows the reconstruction results on three light fields, \textit{Castle}, \textit{Holiday} and \textit{Flowers}, from the CIVIT Dataset~\cite{ICME2018} with upsampling scale $16\times$ (disparity range about 14px). The first case and the third case have thin structures with complex occlusions. The depth and learning-based approach by Kalantari~\textit{et al.}~\cite{DoubleCNN} fails to estimate depth maps accurate enough to warp the input images, and the color CNN cannot correct the misaligned views, producing ghosting artifacts as shown in Fig. \ref{fig:Result2}. For the second case, we demonstrate reconstructed EPIs in a highly non-Lambertian region. Caused by the depth ambiguity, the approach by Kalantari~\textit{et al.}~\cite{DoubleCNN} produces choppiness artifacts along the angular dimension. Due to the limited receptive field of the networks, the results by Wu~\textit{et al.}~\cite{WuEPICNN2018} and Yeung et al.~\cite{YeungECCV2018} show aliasing effects in various degrees. In the demonstrated cases, LLFF~\cite{mildenhall2019local} assigns wrong planes to the tiny structures, leading to ghosting and tearing artifacts. 

Table \ref{table:Result3} and Table \ref{table:Result2} list the quantitative measurements ($8\times$ and $16\times$ upsampling scales) on the light fields from the MPI Light Field Archive~\cite{kiran2017towards} and the CIVIT Dataset~\cite{ICME2018}, respectively. Compared with the baseline approaches, the proposed SAA-Net shows superior performance on both light field datasets.

\begin{table*}
\caption{Quantitative results (PSNR/SSIM) of reconstructed light fields on the light fields from the CIVIT Dataset~\cite{ICME2018}.}
\label{table:Result2}
\vspace{-4mm}
\begin{center}
\begin{tabular}{p{2.5cm}|c|p{1.9cm}<{\centering} p{1.9cm}<{\centering} p{1.9cm}<{\centering} p{1.9cm}<{\centering} p{1.9cm}<{\centering} p{1.9cm}<{\centering}}
& Scale & \textit{Seal \& Balls} & \textit{Castle} & \textit{Holiday} & \textit{Dragon} & \textit{Flowers} & Average\\
\hline
Kalantari~\textit{et al.}~\cite{DoubleCNN}& \multirow{7}*{$8\times$} & 46.83 / 0.990 & 39.14 / 0.973 & 36.03 / 0.979 & 43.97 / 0.989 & 39.00 / 0.989 & 40.99 / 0.984\\
Wu~\textit{et al.}~\cite{WuEPICNN2018}& & 49.01 / 0.997 & 37.67 / 0.984 & 40.46 / 0.995 & 48.38 / 0.997 & 45.85 / 0.998 & 44.27 / 0.994\\
Yeung~\textit{et al.}~\cite{YeungECCV2018} & & 49.83 / 0.997 & 40.84 / 0.993 & 41.16 / 0.996 & 48.61 / 0.997 & 47.83 / 0.997 & 45.65 / 0.996\\
LLFF~\cite{mildenhall2019local} & & 47.03 / 0.990 & 40.25 / 0.988 & 39.69 / 0.987 & 47.38 / 0.990 & 44.71 / 0.991 & 43.81 / 0.989\\
\wgc{HDDRNet~\cite{meng2019high}}&& 46.79 / 0.997 & 40.43 / 0.993 & 39.97 / 0.996 & 47.72 / 0.997 & 43.76 / 0.997 & 43.73 / 0.996\\
\wgc{DA$^2$N~\cite{wu2021revisiting}} &&48.23 / 0.997 & 42.65 / 0.993 & 41.38 / 0.996 & 48.75 / 0.998 & 46.40 / 0.998 & 45.48 / 0.996\\
Our proposed & &\textbf{50.99 / 0.998} & \textbf{43.20 / 0.994} & \textbf{42.29 / 0.997} &\textbf{50.12 / 0.998} &\textbf{48.49 / 0.998} &\textbf{47.02 / 0.997}\\
\hline
Kalantari~\textit{et al.}~\cite{DoubleCNN}& \multirow{12}*{$16\times$} & 43.13 / 0.985 & 36.03 / 0.965 & 32.44 / 0.961 & 39.50 / 0.985 & 35.21 / 0.973 & 37.26 / 0.974\\
Wu~\textit{et al.}~\cite{WuEPICNN2018}& & 45.21 / 0.994 & 35.20 / 0.977 & 35.58 / 0.987 & 46.39 / 0.997 & 41.60 / 0.995 & 40.80 / 0.990\\
Yeung~\textit{et al.}~\cite{YeungECCV2018} & & 44.38 / 0.992 & 37.86 / 0.989 & 36.06 / 0.988 & 45.52 / 0.997 & 42.30 / 0.994 & 41.22 / 0.992\\
LLFF~\cite{mildenhall2019local} & & 45.50 / 0.990 & 38.60 / 0.971 & 36.69 / 0.984 & 44.80 / 0.992 & 41.19 / 0.989 & 41.36 / 0.985\\
\wgc{HDDRNet~\cite{meng2019high}}&& 44.24 / 0.997 & 39.88 / 0.991 & 38.09 / 0.992 & 44.26 / 0.997 & 42.04 / 0.996 & 41.70 / 0.995\\
\wgc{DA$^2$N~\cite{wu2021revisiting}}&& 46.19 / 0.996 & 40.77 / 0.992 & 37.99 / 0.992 & 47.19 / 0.998 & 41.95 / 0.996 & 42.82 / 0.995\\
Our proposed & &\textbf{49.19 / 0.998} & \textbf{41.32 / 0.992} & \textbf{38.88 / 0.993} & \textbf{48.39 / 0.998}  & \textbf{44.05 / 0.997} & \textbf{44.37 / 0.996}\\
\cline{1-1}\cline{3-8}
\wgc{MSR structure-2} & & 46.85 / 0.995 & 37.78 / 0.989 & 36.17 / 0.988 & 47.10 / 0.998 & 42.98 / 0.996 & 42.18 / 0.993\\
\wgc{MSR structure-3} & & 48.50 / 0.997 & 40.66 / 0.991 & 38.23 / 0.992 & 46.94 / 0.997 & 42.92 / 0.996 & 43.45 / 0.995\\
\wgc{MSR structure-4} & & 47.15 / 0.997 & 40.86 / 0.992 & 38.43 / 0.993 & 46.69 / 0.997 & 43.18 / 0.997 & 43.26 / 0.995\\
w/o MSR structure & & 46.41 / 0.994 & 38.65 / 0.990 & 36.78 / 0.988 & 46.83 / 0.997 & 42.77 / 0.996 & 42.29 / 0.993\\
w/o SAP loss & & 48.83 / 0.996 & 41.05 / 0.992 & 38.62 / 0.992 & 48.00 / 0.997  & 43.85 / 0.997 & 44.07 / 0.995\\
\end{tabular}
\end{center}
\vspace{-4mm}
\end{table*}

\subsection{Evaluations on Light Fields from Lytro Illum}
We evaluate the proposed approach using three Lytro light field datasets (113 light fields in total), the \textit{30 Scenes} dataset by Kalantari~\textit{et al.}~\cite{DoubleCNN}, and the \textit{Reflective} and \textit{Occlusions} categories from the Stanford Lytro Light Field Archive~\cite{StanfordLytro}. In this experiment, we reconstruct a $7\times7$ light field from $3\times3$ views ($3\times$ upsampling) and a $8\times8$ light field from $2\times2$ views ($7\times$ upsampling). \wgc{We compare our SAA-Net with 8 learning-based approaches, including 3 depth-based approaches (Kalantari~\textit{et al.}~\cite{DoubleCNN}, LLFF~\cite{mildenhall2019local} and Meng~\textit{et al.}~\cite{meng2021light}) and 5 approaches without explicit depth (Wu~\textit{et al.}~\cite{WuEPICNN2018}, Wang~\textit{et al.}~\cite{wang2020high}, Yeung~\textit{et al.}~\cite{YeungECCV2018}, HDDRNet~\cite{meng2019high} and DA$^2$N~\cite{wu2021revisiting}).} Since the vanilla versions of the networks in~\cite{DoubleCNN,YeungECCV2018,wang2020high,meng2019high,meng2021light} are trained on Lytro light fields, we use their open-source model (parameters) without re-training. \wgc{The networks by Wu~\textit{et al.}~\cite{WuEPICNN2018} and Mildenhall~\textit{et al.}~\cite{mildenhall2019local} (LLFF) are re-trained using the same dataset introduced in Sec. \ref{Sec:train_data}.} Note that the proposed network is not fine-tuned on any Lytro light field datasets, and the results are produced by the same set of network parameters for both $3\times$ and $7\times$ upsampling scales.

Table~\ref{table:Result1} lists the quantitative results on the evaluated Lytro light fields. The proposed SAA-Net shows competitive performance compared with the state-of-the-art light field reconstruction approach by Yeung~\textit{et al.}~\cite{YeungECCV2018}. \wgc{For the evaluated 113 light fields, the average PSNR / SSIM values of the proposed network are 42.55 / 0.984 for $3\times$ upsampling and 36.55 / 0.969 for $7\times$ upsampling. In comparison, the average PSNR / SSIM values of the baseline approaches with the highest performance are 41.82 / 0.984 (DA$^2$N~\cite{wu2021revisiting}) for $3\times$ upsampling and 35.81 / 0.938 (Meng~\textit{et al.}~\cite{meng2021light}) for $7\times$ upsampling.} Since our network is not re-trained or fine-tuned on any Lytro light field dataset, these experimental results clearly demonstrate that our network can generalize well on light fields captured by different acquisition geometries.

\begin{figure*}
	\begin{center}
		\includegraphics[width=1\linewidth]{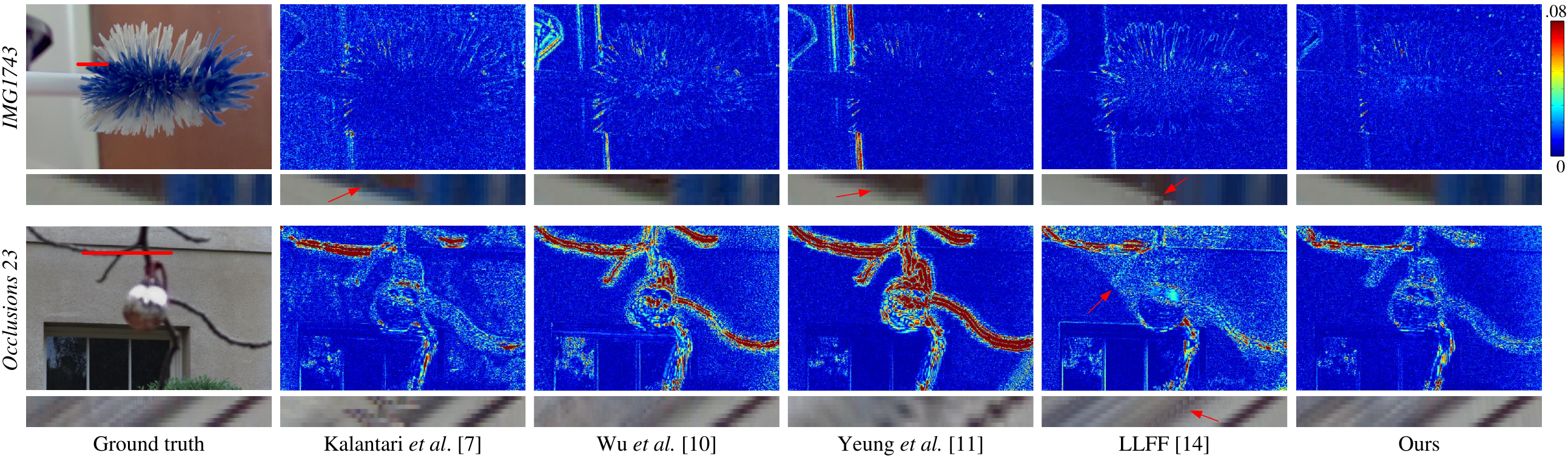}
	\end{center}
	\vspace{-4mm}
	\caption{Comparison of the results on the light fields from Lytro Illum. The results show the error map (absolute error of the grey-scale image) and the EPIs at the location marked by red lines. Light fields are from the \textit{30 Scenes}~\cite{DoubleCNN} and the \textit{Occlusions} category~\cite{StanfordLytro}. Please zoom-in for better visual comparison.}
	\label{fig:Result1}
\end{figure*}

We demonstrate two cases with relatively large disparities (maximum disparity up to 13px), \textit{IMG1743} from the \textit{30 Scenes}~\cite{DoubleCNN} and \textit{Occlusions 23} from the \textit{Occlusions} category~\cite{StanfordLytro}, as shown in Fig. \ref{fig:Result1}. In both cases, the reconstruction results by Wu~\textit{et al.}~\cite{WuEPICNN2018} and Yeung et al.~\cite{YeungECCV2018} show ghosting artifacts around the region with large disparity (background in the \textit{IMG1743} case, and foreground in the \textit{Occlusions 23} case) due to their limited receptive fields. \wgc{The depth and learning-based approaches by Kalantari~\textit{et al.}~\cite{DoubleCNN} and Mildenhall~\textit{et al.}~\cite{mildenhall2019local} (LLFF) produce promising results in the first case, but appear tearing artifacts near the occlusion boundaries as marked by the red arrows in the EPIs.} In the second case, the approach by Kalantari~\textit{et al.}~\cite{DoubleCNN} fails to estimate proper depth information, introducing misalignment as shown by the EPI. \wgc{LLFF~\cite{mildenhall2019local} produces ghosting effects due to the incorrect plane assignments around the background region, as shown by the red arrows in the figure.} In comparison, the proposed SAA-Net provides reconstructed light fields with higher view consistency (as shown in the demonstrated EPIs).

\begin{table}
\caption{Quantitative results (PSNR/SSIM) of reconstructed views on the light fields from Lytro Illum~\cite{Lytro}. The \textit{30 Scenes} dataset courtesy of Kalantari~\textit{et al.}~\cite{DoubleCNN}, and the \textit{Reflective} (32 light fields) and \textit{Occlusions} (51 light fields) categories are from the Stanford Lytro Light Field Archive~\cite{StanfordLytro}.}
\label{table:Result1}
\vspace{-3mm}
\begin{center}
\begin{tabular}{l|c|ccc}
& Scale & \textit{30 Scenes}  &\textit{Reflective} &\textit{Occlusions}\\
\hline
\scriptsize{Kalantari~\textit{et al.}}~\cite{DoubleCNN}& \multirow{8}*{$3\times$} & 39.62/0.978 & 37.78/0.971 & 34.02/0.955 \\
Wu~\textit{et al.}~\cite{WuEPICNN2018}& & 41.02/0.988 & 41.71/0.989 & 38.11/0.944 \\
Wang~\textit{et al.}~\cite{wang2020high}& & 43.82/0.993 & 39.93/0.959  &34.69/0.923\\
Yeung~\textit{et al.}~\cite{YeungECCV2018}& & 44.53/0.990 & 42.56/0.975  &39.27/0.945\\
\wgc{LLFF~\cite{mildenhall2019local}} & & 39.92/0.977 & 39.52/0.969 & 35.64/0.929 \\
\wgc{HDDRNet~\cite{meng2019high}}& & 43.02/0.989 & 40.72/0.979  & 36.13/0.958\\
\wgc{DA$^2$N~\cite{wu2021revisiting}}&&43.69/0.995&43.25/0.991 & 39.82/0.971  \\
Our proposed & & \textbf{44.75}/\textbf{0.996} & \textbf{44.04}/\textbf{0.992} & \textbf{40.32}/\textbf{0.972} \\
\hline
\scriptsize{Kalantari~\textit{et al.}}~\cite{DoubleCNN}& \multirow{13}*{$7\times$} & 38.21/0.974 & 35.84/0.942 & 31.81/0.895 \\
Wu~\textit{et al.}~\cite{WuEPICNN2018}& & 36.28/0.965 & 36.48/0.962 & 32.19/0.907 \\
Yeung~\textit{et al.}~\cite{YeungECCV2018}& & 39.22/0.977 & 36.47/0.947  & 32.68/0.906\\
\wgc{LLFF~\cite{mildenhall2019local}} & & 38.17/0.974 & 36.40/0.948 & 31.96/0.901 \\
\wgc{HDDRNet~\cite{meng2019high}}& & 38.33/0.967 & 36.77/0.931  & 32.78/0.909\\
\wgc{Meng~\textit{et al.}~\cite{meng2021light}}& & 39.14/0.970 & 37.01/0.950  & 33.10/0.912\\
\wgc{DA$^2$N~\cite{wu2021revisiting}}   &        & 38.99/0.986 & 36.72/0.975 & 33.14/0.950 \\
Our proposed & & \textbf{39.98}/\textbf{0.988} & \textbf{37.77}/\textbf{0.978} & \textbf{33.77}/\textbf{0.952} \\
\cline{1-1}\cline{3-5}
\wgc{MSR structure-2} & & 37.17/0.978 & 36.29/0.974 & 32.02/0.944\\
\wgc{MSR structure-3} & & 37.62/0.980 & 36.93/0.975 & 32.53/0.948\\
\wgc{MSR structure-4} & & 37.20/0.981 & 36.67/0.975 & 32.58/0.947\\
\wgc{\scriptsize{w/o MSR structure}} & & 37.08/0.978 & 36.37/0.975 & 32.37/0.946 \\
\wgc{w/o SAP loss} & & 39.71/0.986 & 37.39/0.977 & 33.52/0.950 \\
\end{tabular}
\end{center}
\vspace{-4mm}
\end{table}

\subsection{Ablation studies}
In this experiment, we empirically validate the modules and losses in our SAA-Net by performing the following ablation studies \wgc{on different datasets. The results are listed in the last five rows in Table \ref{table:Result2}, \ref{table:Result3} and \ref{table:Result1}. First, we evaluate the network with increasing series of the MSR structure while removing the SAAM. We use ``MSR structure-$X$'' to represent the network with $X$ series of MSR structure. By increasing the series, the network will have a larger receptive field, e.g., the theoretical receptive field size of the MSR structure-4 reaches 247 pixels in the spatial dimension. However, since it is intractable to increase the actual size of the receptive field simply by using a deeper network~\cite{zhou2015object}, purely increasing the series of MSR structure has limitations in improving performance. This point is also verified by the ablation study. In comparison, the non-local attention is more effective than simply increasing the receptive field of the network.}

In the second ablation study, we use a typical 3D U-net with the same convolutional layers as the backbone and remove the deconvolution layer in the skip connections, denoted as ``w/o MSR structure'' for short. The angular reconstruction is simply achieved by using deconvolution at the end of the network. The network parameters are kept comparable to the SAA-Net by adjusting the channel dimension of the network. \wgc{For light fields from gantry systems~\cite{kiran2017towards,ICME2018}, the performance of the network decreases more than 2dB in terms of PSNR, as shown in Table \ref{table:Result2} and \ref{table:Result3}. For light fields from Lytro Illum, the performance of the network decreases more than 1.3dB, as shown in Table \ref{table:Result1}.} 

In the last ablation study, we train the proposed SAA-Net simply by using the pixel-wise term (MAE loss) without the proposed spatial-angular perceptual loss, denoted as ``w/o SAP loss'' for short. The performance (PSNR) decreases around 0.3dB as shown in Table \ref{table:Result2}, \ref{table:Result3} and \ref{table:Result1}.

\begin{figure*}
\begin{center}
\includegraphics[width=1\linewidth]{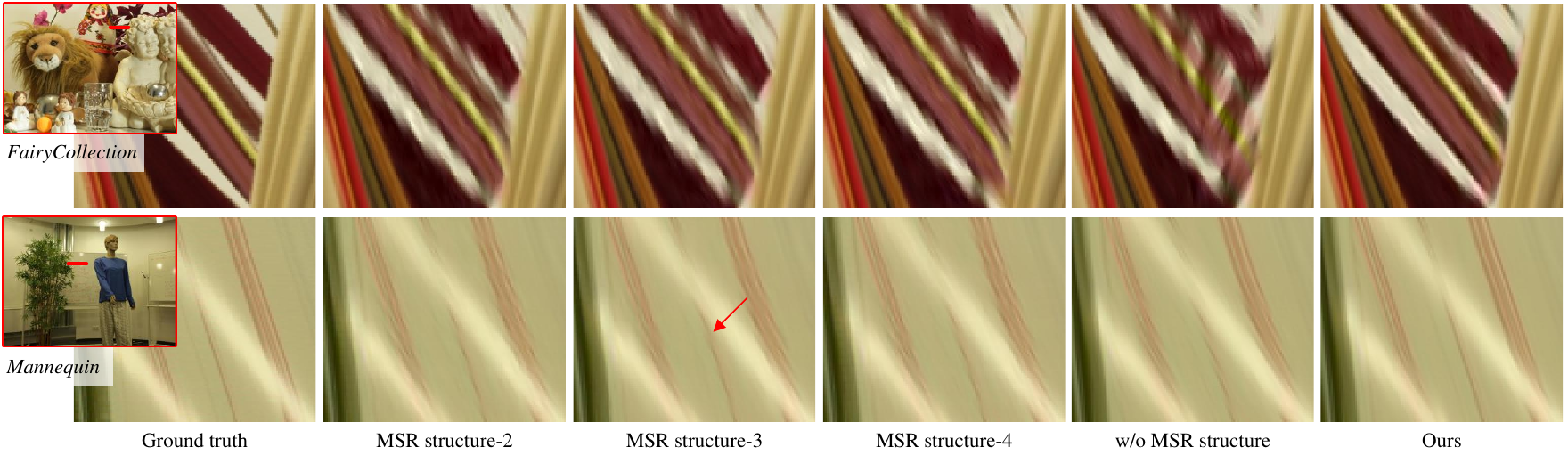}
\end{center}
\vspace{-4mm}
   \caption{ We compare our SAA-Net against the network with different series of the MSR structure without using the SAAM (MSR structure-$X$) and the network without the MSR structure on the light fields from the MPI Light Field Archive~\cite{kiran2017towards} ($16\times$ upsampling).}
\label{fig:ablation}
\end{figure*}

\begin{figure*}
\begin{center}
\includegraphics[width=1\linewidth]{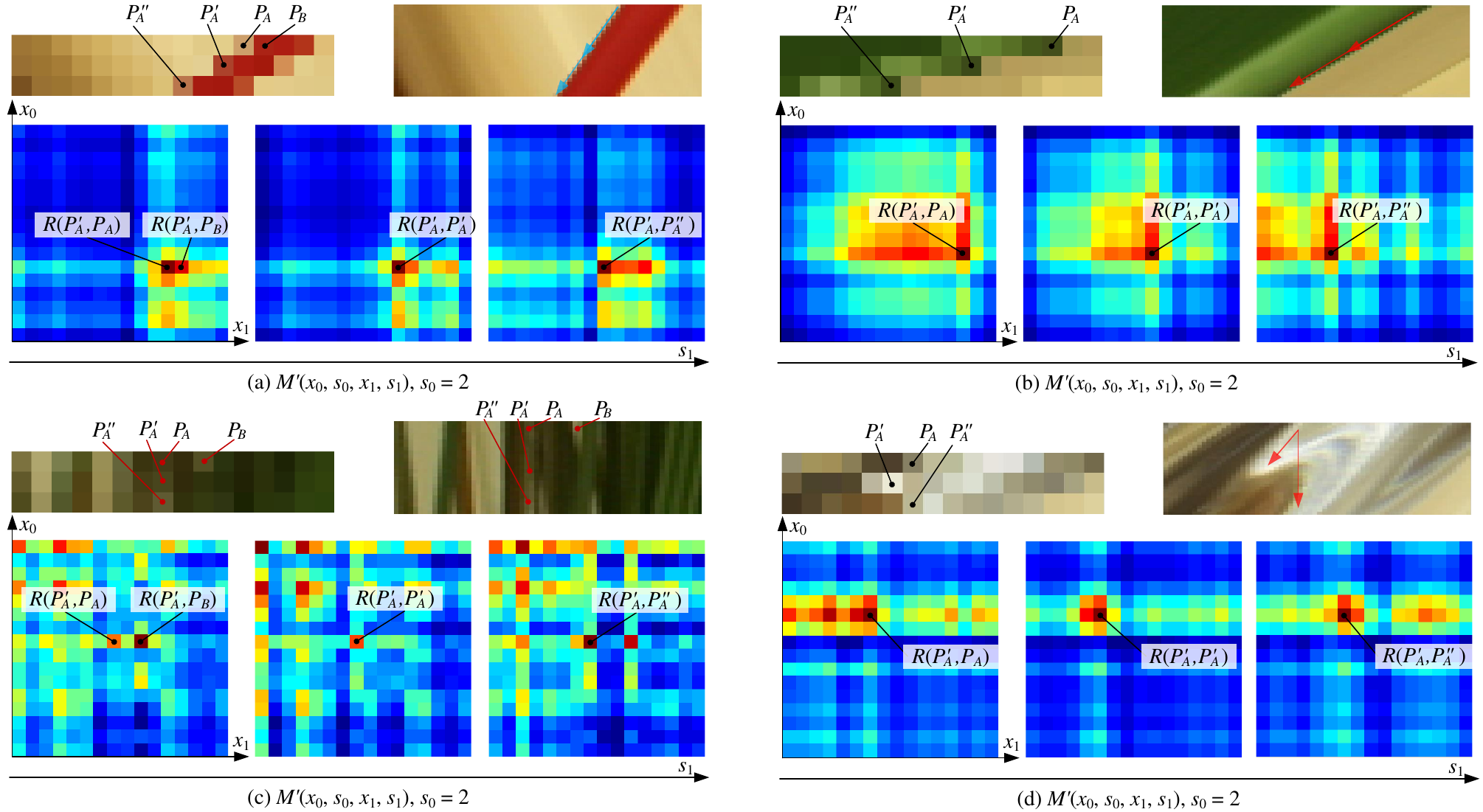}
\end{center}
\vspace{-4mm}
   \caption{Additional results of attention map (before the softmax function) on scenes with \wgc{(a) small disparity, (b) large disparity, (c) occlusion and (b) non-Lambertian effect. In each case, the top left figure shows the input EPI, the top right figure shows the reconstructed EPI, and the bottom figure visuals three sub-maps.}}
\label{fig:att_map1}
\end{figure*}

\wgc{In Fig.~\ref{fig:ablation}, we also visualize the comparison ($16\times$ upsampling) between the proposed SAA-Net, the network with different series of the MSR structure while removing the SAAM and the network without the MSR structure. The evaluated light fields are from the MPI Light Field Archive [16]. The results show that without using the proposed modules, especially the MSR structure, the SAA-Net appears severe aliasing effects around region with large disparity, e.g., the occluded leaves in the \textit{FairyCollection} case and the hand writing on the reflective board in the \textit{Mannequin} case.}

\section{Further Analysis}
\subsection{Spatial-Angular Attention Map}\label{Sec:attention_map}
We visualize four additional attention maps (before the softmax function) on scenes with small disparity, large disparity, occlusion and non-Lambertian effect, as shown in Fig. \ref{fig:att_map1}. \wgc{The first case (Fig. \ref{fig:att_map1}(a)) shows a scene with a relatively small disparity (about 7 pixels). Due to the spatial downsampling operation in the MSR structure, the disparity of the light field features in the SAAM is about 1.75 pixels, as shown in the top left figure in Fig. \ref{fig:att_map1}(a). Three sub-maps $M'(x_0,s_0,x_1,s_1), s_0=2, s_1=1,2,3$ are shown in the bottom of Fig. \ref{fig:att_map1}(a), which visualize the correspondence captured by the attention mechanism. As we can see, the response with the highest value moves from $R(P_A',P_A)$ at position $M'(11,2,12,1)$ to $R(P_A',P_A'')$ at position $M'(11,2,9,3)$ along the angular dimension. In addition, the attention map also shows a high response value $R(P_A',P_B)$ at position $M'(11,2,13,1)$ due to the fractional disparity. This indicates that the proposed SAAM has potential to capture a correspondence with sub-pixel accuracy.}

In the second case (Fig. \ref{fig:att_map1}(b)), we demonstrate the spatial-angular attention on a scene with a large disparity (about 16 pixels). The spatial downsampling operation in the MSR structure reduces the disparity to 4 pixels. The response in the attention map moves from $R(P_A',P_A)$ at position $M'(10,2,14,1)$ to $R(P_A',P_A'')$ at position $M'(10,2,6,3)$ along the angular dimension. \wgc{Although the SAAM only applies $1\times1\times1$ convolutions before the attention, it is capable to capture correspondence with large displacement.}

\wgc{Fig. \ref{fig:att_map1}(c) visualize the attention map of a scene with a complex occlusion structure (\textit{Mannequin} in the MPI Light Field Archive~\cite{kiran2017towards}). The background white board is occluded by the foreground leaves (please refer to the sub-aperture image of the second row in Fig. \ref{fig:ablation}). As shown by the demonstrated attention map in Fig. \ref{fig:att_map1}(c), the SAAM fails to capture a correct correspondence because of the occlusion. The highest response value is $R(P_A',P_B)$ at position $M'(8,2,10,1)$, which is also a background point like $P_A$. However, the SAAM also generate a considerable response $R(P_A',P_A)$ at position $M'(8,2,8,1)$, which we speculate is inferred from other non-occluded background points.}

The fourth case in Fig. \ref{fig:att_map1}(d) demonstrates the spatial-angular attention on a scene with non-Lambertian effect. In this case, the positional relation between the corresponding points $P_A$, $P_A'$ and $P_A''$ does not follow \wgc{a clear depth cue}, as clearly shown in the top right figure of Fig. \ref{fig:att_map1}(d). \wgc{Analogously, the responses do not follow a regular disparity pattern along the angular dimension, as visualized by the at the bottom of Fig. \ref{fig:att_map1}(d).} This result shows that the proposed SAAM is able to catch the correspondences even for regions with non-Lambertian effects.

\subsection{Tensor Decomposition for Spatial-Angular Attention}
Although we propose a multi-scale reconstruction structure to save GPU memory cost, the SAAM will still occupy a large GPU memory when dealing with an input light field with high spatial-angular resolution. For example, when reconstructing light fields from the MPI Light Field Archive~\cite{kiran2017towards} (spatial resolution $960\times720$), we have to disassemble the 3D data into sub-light fields of resolution $960\times 24\times25$ (width $\times$ height $\times$ angular). Our investigation shows that the disassembling will cause quality degradation on the reconstructed results.

\begin{figure}
	\begin{center}
		\includegraphics[width=1.\linewidth]{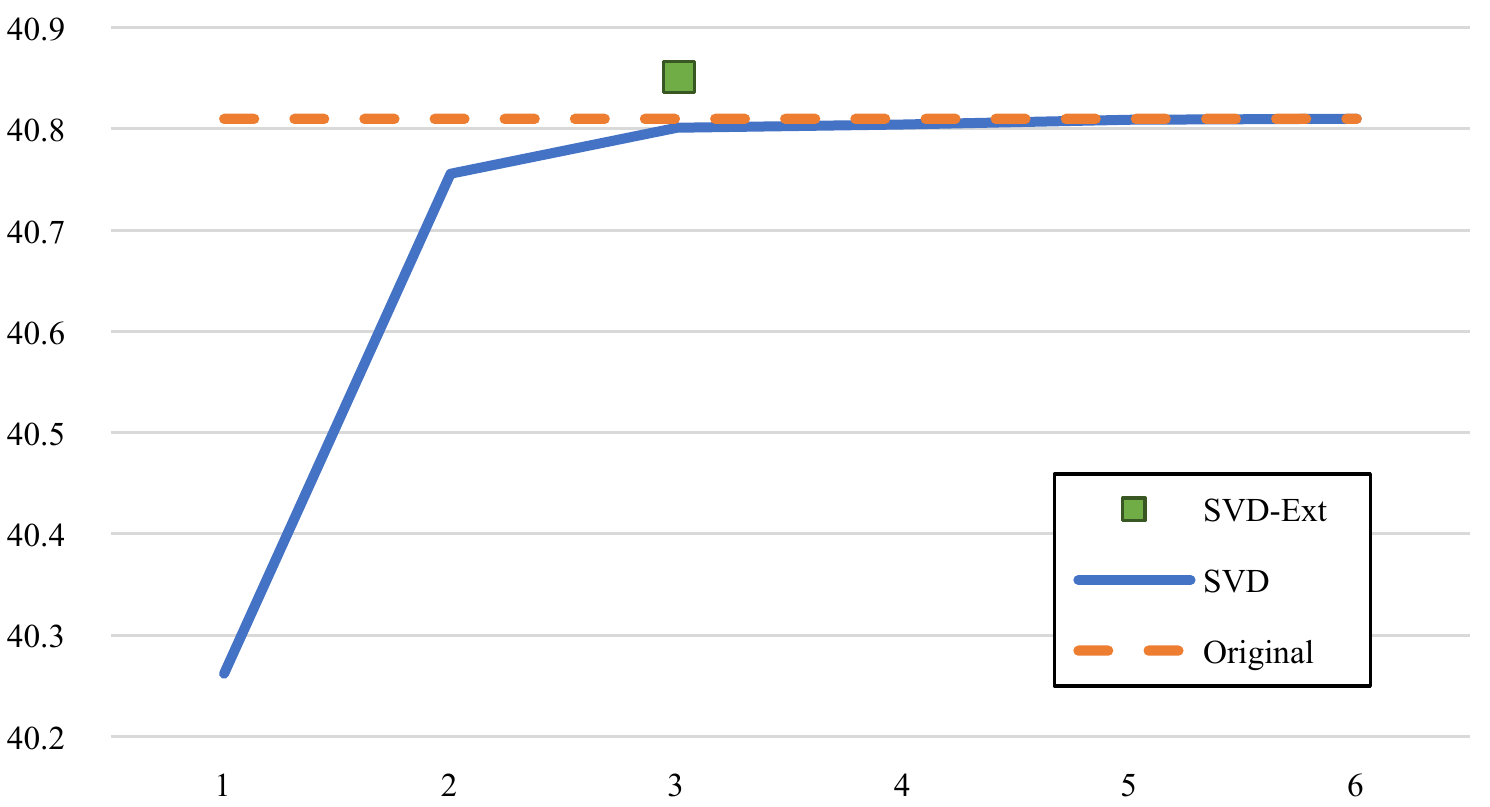}
	\end{center}
	\vspace{-4mm}
	\caption{The performance curve (PSNR) against the SVD decomposition of the proposed SAAM. The ``SVD'' denotes the truncated SVD with different parameters $\tau$. The ``Original'' denotes the SAAM without SVD decomposition. The ``SVD-Ext'' denotes the truncated SVD with parameter $\tau=3$ and the input sub-light fields of resolution $960\times 64\times25$. The results are averaged on the 5 light fields from the MPI Light Field Archive~\cite{kiran2017towards}.}
	\label{fig:SVD}
\end{figure}

We therefore apply the truncated Singular Value Decomposition (SVD)~\cite{girshick2015fast} to compact the 3D tensor $\phi'_q\in\mathbb{R}^{BH\times WA\times C'}$ and $\phi'_k\in\mathbb{R}^{BH\times C'\times WA}$ before computing the attention map
$$\tilde{\phi}=USV^T,$$
where $\tilde{\phi}$ denotes $\phi'_q$ or $\phi^{'T}_k$, $U$ and $V$ are two orthogonal matrices (ignoring the batch dimension), and $S$ is a diagonal matrix with singular values along its diagonal. By truncating the diagonal matrix $S$ with the largest $\tau$ singular values, we can get a good approximation $\tilde{\phi}\approx US_{\tau}V^T$ and also compress the 3D tensors. Since the rank of the matrices are $C'=6$, the parameter of the truncated SVD $\tau=[1,2,\cdots,C']$.

Fig.~\ref{fig:SVD} shows the performance on PSNR in function of the SVD decomposition using the largest $\tau=[1,2,\cdots,6]$ singular values. As we can see from the ``SVD'' curve, with no less than three singular values, the SVD decomposition will maintain the network performance without using fine-tuning. Moreover, since the decomposition enables us to feed the network with higher spatial resolution input, e.g., from $960\times 24\times25$ to $960\times 64\times25$, we can obtain a reconstruction result with even higher quality (0.04dB higher) when employing truncated SVD decomposition, as shown by the ``SVD-Ext'' (green dot) in the figure.

\subsection{Limitations}
The non-local attention involves outer product of large scale matrices, especially for the high-dimensional light field data. For this reason, the proposed network takes almost 15\% of its inference time on the SAAM. For the 3D light field from the MPI Light Field Archive~\cite{kiran2017towards}, the network takes about 51 seconds to reconstruct a $1\times97$ light field from $1\times7$ views of spatial resolution $960\times720$ ($16\times$ upsampling), i.e., 0.53s per view. For the 3D light field from the CIVIT Dataset~\cite{ICME2018}, the network takes about 126 seconds to reconstruct a $1\times193$ light field from $1\times13$ views of spatial resolution $1280\times720$ ($16\times$ upsampling), i.e., 0.65s per view. For a 4D light field from Lytro Illum, it takes about 17 seconds to reconstruct a $7\times7$ light field from $3\times3$ views of spatial resolution $536\times376$ ($3\times$ upsampling), i.e., about 0.35s per view. And the reconstruction of a $8\times8$ Lytro light field from $2\times2$ views ($7\times$ upsampling) takes about 30 seconds, i.e., about 0.5s per view. The parameter number of our SAA-Net is about 338K. The above evaluations are performed on an Intel Xeon Gold 6130 CPU @ 2.10GHz with an NVIDIA TITAN Xp.

Although we apply a simple SVD decomposition to accerlate the network and compact the 3D tensor, the compression rate is limited by the rank of the matrices. Decomposing the attention map into the combination of small tensors~\cite{kolda2009tensor} might solve this problem in a more essential way.

The another limitation of our proposed method is that repetitive patterns in the input light field can cause multiple plausible responses in the attention map, leading to misalignments in the reconstructed light fields. A possible solution is to introduce a smooth term in the attention map as in~\cite{wang2019learning} to penalize multiple responses. 

\section{Conclusions}
We have proposed a spatial-angular attention module in a 3D U-net backbone to capture correspondence information non-locally for light field reconstruction. The introduced Spatial-Angular Attention Module (termed as SAAM) is designed to compute the responses from all the positions on the epipolar plane for each pixel in the light field and produce a spatial-angular attention map that records the correspondences. The attention map is then applied to guide light field reconstruction via channel-to-angular pixel shuffling. We further propose a multi-scale reconstruction structure based on the 3D U-net backbone that implements the SAAM efficiently in the low spatial resolution feature space, while also preserving fine details in the high spatial resolution feature space. For the network training, a spatial-angular perceptual loss is designed specifically for the high-dimensional light field data by pre-training a 3D auto-encoder. The evaluations on light fields with challenging non-Lambertian effects and large disparities have demonstrated the superiority of the proposed spatial-angular attention network.


%

%
%

\ifCLASSOPTIONcaptionsoff
  \newpage
\fi



\bibliographystyle{IEEEtran}
\bibliography{IEEEabrv}
%
%
%

%

\begin{IEEEbiography}[{\includegraphics[width=1in]{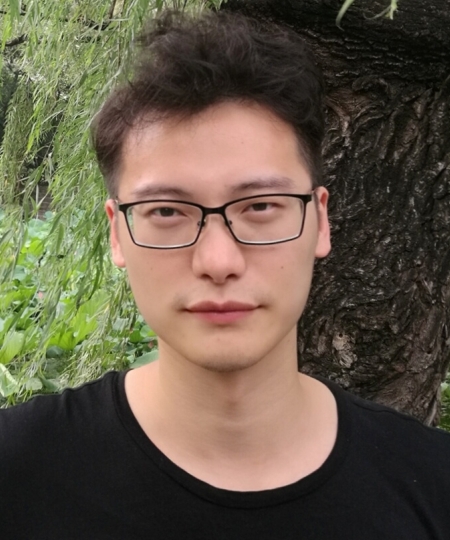}}]{Gaochang Wu}
received the BE and MS degrees in mechanical engineering in Northeastern University, Shenyang, China, in 2013 and 2015, respectively, and Ph.D. degree in control theory and control engineering in Northeastern University, Shenyang, China in 2020. He is currently an associate professor in the State Key Laboratory of Synthetical Automation for Process Industries, Northeastern University. His current research interests include image processing, light field processing and deep learning.
\end{IEEEbiography}

\begin{IEEEbiography}[{\includegraphics[width=1in]{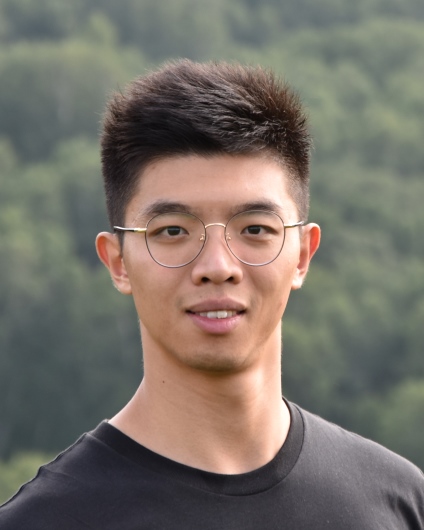}}]{Yingqian Wang}
received the B.E. degree in electrical engineering from Shandong University (SDU), Jinan, China, in 2016, and the M.E. degree in information and communication engineering from National University of Defense Technology (NUDT), Changsha, China, in 2018. He is currently pursuing the Ph.D. degree with the College of Electronic Science and Technology, NUDT. His research interests focus on low-level vision, particularly on light field imaging and image super-resolution.
\end{IEEEbiography}

\begin{IEEEbiography}[{\includegraphics[width=1in]{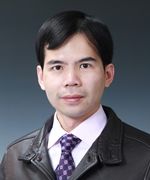}}]{Yebin Liu}
received the BE degree from Beijing University of Posts and Telecommunications, China, in 2002, and the PhD degree from the Automation Department, Tsinghua University, Beijing, China, in 2009. He has been working as a research fellow at the computer graphics group of the Max Planck Institute for Informatik, Germany, in 2010. He is currently an associate professor in Tsinghua University. His research areas include computer vision and computer graphics.
\end{IEEEbiography}

\begin{IEEEbiography}[{\includegraphics[width=1in]{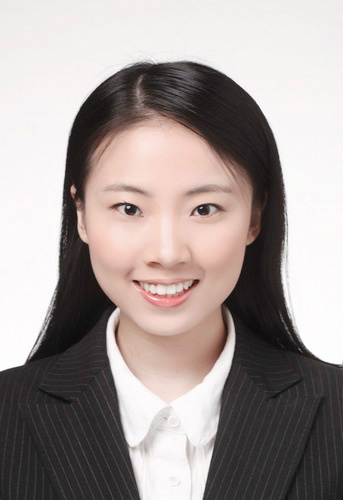}}]{Lu FANG}
is currently an Associate Professor at Tsinghua University. She received her Ph.D in Electronic and Computer Engineering from HKUST in 2011, and B.E. from USTC in 2007, respectively. Dr. Fang's research interests include image / video processing, vision for intelligent robot, and computational photography. Dr. Fang serves as TC member in Multimedia Signal Processing Technical Committee (MMSP-TC) in IEEE Signal Processing Society.
\end{IEEEbiography}

\begin{IEEEbiography}[{\includegraphics[width=1in]{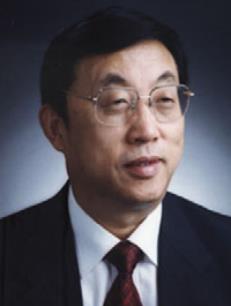}}]{Tianyou Chai}
received the Ph.D. degree in control theory and engineering from Northeastern University, Shenyang, China, in 1985. He has been with the Research Center of Automation, Northeastern University, Shenyang, China, since 1985, where he became a Professor in 1988 and a Chair Professor in 2004. His current research interests include adaptive control, intelligent decoupling control, integrated plant control and systems, and the development of control technologies with applications to various industrial processes. Prof. Chai is a member of the Chinese Academy of Engineering, an academician of International Eurasian Academy of Sciences, IEEE Fellow and IFAC Fellow. He is a distinguished visiting fellow of The Royal Academy of Engineering (UK) and an Invitation Fellow of Japan Society for the Promotion of Science (JSPS).
\end{IEEEbiography}




\end{document}